\documentclass[twocolumn, times, tigten, appendixfloats]{aastex7}

\usepackage{amsmath}
\usepackage{romannum}
\usepackage{natbib}
\usepackage{enumitem}
\usepackage{soul}
\usepackage{longtable}

\usepackage{graphicx}           
\usepackage{latexsym}           
\usepackage{bm}                 
\usepackage[english]{babel}
\usepackage{color}              

\newcommand{\Msol}{\hbox{M$_\sun$}}

\newcommand{\no}{\nodata}

\begin{document}

\pagenumbering{arabic}

\title{PEARLS: 21 Transients Found in the Three-Epoch NIRCam Observations in the Continuous Viewing Zone of the James Webb Space Telescope}

\author[0000-0001-7592-7714]{Haojing Yan}
\affiliation{Department of Physics and Astronomy, University of Missouri, Columbia, MO 65211, USA}
\email{yanha@missouri.edu}

\author[0000-0001-7957-6202]{Bangzheng Sun}
\affiliation{Department of Physics and Astronomy, University of Missouri, Columbia, MO 65211, USA}
\email{bangzheng.sun@mail.missouri.edu}

\author[0000-0003-3270-6844]{Zhiyuan Ma}
\affiliation{Department of Astronomy, University of Massachusetts, Amherst, MA 01003, USA}
\email{zhiyuanma@umass.edu}

\author[0000-0001-7092-9374]{Lifan Wang}
\affiliation{George P. and Cynthia Woods Mitchell Institute for Fundamental Physics \& Astronomy, \\
Texas A. \& M. University, Department of Physics and Astronomy, 4242 TAMU, College Station, TX 77843, USA}
\email{lifan@tamu.edu}

\author[0000-0001-9262-9997]{Christopher N. A. Willmer} 
\affiliation{Steward Observatory, University of Arizona,
933 N Cherry Ave, Tucson, AZ, 85721-0009, USA}
\email{cnaw@arizona.edu}

\author[0000-0003-1060-0723]{Wenlei Chen}
\affiliation{Department of Physics, Oklahoma State University, 145 Physical Sciences Bldg, Stillwater, OK 74078, USA}
\email{wenlei.chen@okstate.edu}

\author[0000-0001-9440-8872]{Norman A. Grogin} 
\affiliation{Space Telescope Science Institute,
3700 San Martin Drive, Baltimore, MD 21218, USA}
\email{nagrogin@stsci.edu}

\author[0000-0002-0005-2631]{John F. Beacom} 
\affiliation{Center for Cosmology and AstroParticle Physics (CCAPP), The Ohio State University, Columbus, OH 43210, USA}
\affiliation{Department of Physics, The Ohio State University, Columbus, OH 43210, USA}
\affiliation{Department of Astronomy, The Ohio State University, Columbus, OH 43210, USA}
\email{beacom.7@osu.edu}

\author[0000-0002-9895-5758]{S. P. Willner}
\affiliation{Center for Astrophysics \textbar\ Harvard \& Smithsonian, 60 Garden Street, Cambridge, MA 02138, USA}
\email{swillner@cfa.harvard.edu}

\author[0000-0003-3329-1337]{Seth H.~Cohen} 
\affiliation{School of Earth \& Space Exploration, Arizona State University, Tempe, AZ 85287-1404, USA}
\email{seth.cohen@asu.edu}

\author[0000-0001-8156-6281]{Rogier A.~Windhorst}
\affiliation{School of Earth \& Space Exploration, Arizona State University, Tempe, AZ 85287-1404, USA}
\affiliation{Department of Physics, Arizona State University, Tempe, AZ 85287-1504, USA}
\email{rogier.windhorst@asu.edu}

\author[0000-0003-1268-5230]{Rolf A.~Jansen}
\affiliation{School of Earth \& Space Exploration, Arizona State University, Tempe, AZ 85287-1404, USA}
\email{rolf.jansen@asu.edu}

\author[0000-0003-0202-0534]{Cheng Cheng}
\affiliation{Chinese Academy of Sciences South America Center for Astronomy, National Astronomical Observatories, CAS, Beijing 100101, China}
\email{chengcheng@nao.cas.cn}

\author[0000-0001-6511-8745]{Jia-Sheng Huang}
\affiliation{Chinese Academy of Sciences South America Center for Astronomy, National Astronomical Observatories, CAS, Beijing 100101, People's Republic of China}
\affiliation{Center for Astrophysics \textbar\ Harvard \& Smithsonian, 60 Garden Street, Cambridge, MA 02138, USA}
\email{jhuang@nao.cas.cn}

\author[0000-0001-7095-7543]{Min Yun}
\affiliation{Department of Astronomy, University of Massachusetts, Amherst, MA 01003, USA}
\email{myun@umass.edu}

\author[0000-0003-1436-7658]{Hansung B. Gim}
\affiliation{Department of Physics, Montana State University, Bozeman, MT 59717, USA}
\email{hansung.b.gim@gmail.com} 

\author[0000-0001-8751-3463]{Heidi B.~Hammel} 
\affiliation{Association of Universities for Research in Astronomy, Washington, DC 20004, USA}
\email{hbhammel@aura-astronomy.org}

\author[0000-0001-7694-4129]{Stefanie N.~Milam} 
\affiliation{NASA Goddard Space Flight Center, Greenbelt, MD 20771, USA}
\email{stefanie.n.milam@nasa.gov}

\author[0000-0002-6610-2048]{Anton M.\ Koekemoer}  
\affiliation{Space Telescope Science Institute,
3700 San Martin Drive, Baltimore, MD 21218, USA}
\email{koekemoer@stsci.edu}

\author[0000-0001-7201-1938]{Lei Hu}
\affiliation{McWilliams Center for Cosmology and Astrophysics, 
Department of Physics, Carnegie Mellon University,
5000 Forbes Ave, Pittsburgh, PA 15213, USA}
\email[]{leihu@andrew.cmu.edu}

\author[0000-0001-9065-3926]{Jos\'e M. Diego} 
\affiliation{Instituto de Física de Cantabria (CSIC-UC), 
Avda.\ Los Castros s/n, 39005, Santander, Spain}
\email{jdiego@ifca.unican.es}

\author[0000-0002-7265-7920]{Jake Summers}
\affiliation{School of Earth \& Space Exploration, Arizona State University, Tempe, AZ 85287-1404, USA}
\email{jssumme1@asu.edu}

\author[0000-0002-9816-1931]{Jordan C.\ J.\ D'Silva}  
\affiliation{International Centre for Radio Astronomy Research (ICRAR) and the
International Space Centre (ISC), The University of Western Australia, M468,
35 Stirling Highway, Crawley, WA 6009, Australia}
\affiliation{ARC Centre of Excellence for All Sky Astrophysics in 3 Dimensions
(ASTRO 3D), Australia}
\email{jordan.dsilva@research.uwa.edu.au}

\author[0000-0001-7410-7669]{Dan Coe}  
\affiliation{Space Telescope Science Institute, 3700 San Martin Drive, Baltimore, MD 21218, USA}
\affiliation{Association of Universities for Research in Astronomy (AURA) for the European Space Agency (ESA), STScI, Baltimore, MD 21218, USA}
\affiliation{Center for Astrophysical Sciences, Department of Physics and Astronomy, The Johns Hopkins University, 3400 N Charles St. Baltimore, MD 21218, USA}
\email{dcoe@stsci.edu}

\author[0000-0003-1949-7638]{Christopher J.\ Conselice}
\affiliation{Jodrell Bank Centre for Astrophysics, Alan Turing Building,
University of Manchester, Oxford Road, Manchester M13 9PL, UK}
\email{conselice@gmail.com}

\author[0000-0001-9491-7327]{Simon P.\ Driver}  
\affiliation{International Centre for Radio Astronomy Research (ICRAR) and the
International Space Centre (ISC), The University of Western Australia, M468,
35 Stirling Highway, Crawley, WA 6009, Australia}
\email{Simon.Driver@icrar.org}

\author[0000-0003-1625-8009]{Brenda Frye}
\affiliation{Department of Astronomy/Steward Observatory, University of Arizona, 933 N Cherry Ave, Tucson, AZ, 85721-0009, USA}
\email{bfrye@arizona.edu}

\author[0000-0001-6434-7845]{Madeline A.\ Marshall} 
\affiliation{Los Alamos National Laboratory, Los Alamos, NM 87545, USA}
\email{madeline_marshall@outlook.com}

\author[0000-0002-6150-833X]{Rafael {Ortiz~III}} 
\affiliation{School of Earth \& Space Exploration, Arizona State University, Tempe, AZ 85287-1404, USA}
\email{rortizii@asu.edu}

\author[0000-0003-3382-5941]{Nor Pirzkal}  
\affiliation{Space Telescope Science Institute,
3700 San Martin Drive, Baltimore, MD 21218, USA}
\email{npirzkal@stsci.edu}

\author[0000-0003-0429-3579]{Aaron Robotham}  
\affiliation{International Centre for Radio Astronomy Research (ICRAR) and the
International Space Centre (ISC), The University of Western Australia, M468,
35 Stirling Highway, Crawley, WA 6009, Australia}
\email{aaron.robotham@uwa.edu.au}

\author[0000-0003-0894-1588]{Russell E.\ Ryan, Jr.}  
\affiliation{Space Telescope Science Institute,
3700 San Martin Drive, Baltimore, MD 21218, USA}
\email{rryan@stsci.edu}

\author[0000-0002-9984-4937]{Rachel Honor}  
\affiliation{School of Earth \& Space Exploration, Arizona State University, Tempe, AZ 85287-1404, USA}
\email{rchonor@asu.edu}

\author[0000-0003-3351-0878]{Rosalia O'Brien}
\affiliation{School of Earth \& Space Exploration, Arizona State University, Tempe, AZ 85287-1404, USA}
\email{robrien5@asu.edu}

\author[0000-0002-0670-0708]{Giovanni G. Fazio}
\affiliation{Center for Astrophysics \textbar\ Harvard \& Smithsonian, 60 Garden Street, Cambridge, MA 02138, USA}
\email{gfazio@cfa.harvard.edu}

\author[0000-0003-4875-6272]{Nathan J. Adams} 
\affiliation{Jodrell Bank Centre for Astrophysics, Alan Turing Building,
University of Manchester, Oxford Road, Manchester M13 9PL, UK}
\email{nathan.adams@manchester.ac.uk}

\author[0000-0003-4223-7324]{Massimo Ricotti}
\affiliation{Department of Astronomy, University of Maryland, College Park, 20742, USA}
\email{ricotti@umd.edu}

\author[0000-0002-5319-6620]{Payaswini Saikia}
\affiliation{Center for Astrophysics and Space Science (CASS), New York University Abu Dhabi, PO Box 129188, Abu Dhabi, UAE }
\email{payaswini.ssc@gmail.com}

\author[0000-0001-6145-5090]{Nimish P. Hathi}
\affiliation{Space Telescope Science Institute, 3700 San Martin Drive,
Baltimore, MD 21218, USA}
\email{nphathi@gmail.com}

\author[0000-0002-0648-1699]{Brent Smith}
\affiliation{School of Earth \& Space Exploration, Arizona State University, Tempe, AZ 85287-1404, USA}
\email{bsmith18@asu.edu}

\author[0000-0002-4884-6756]{Benne W. Holwerda}
\affiliation{Department of Physics and Astronomy, University of Louisville, 102 Natural Science Building, Louisville, KY 40292 USA}
\email{bwholw01@louisville.edu}

\author[0000-0003-3142-997X]{Patrick Kelly}
\affiliation{School of Physics and Astronomy, University of Minnesota, 116
Church Street SE, Minneapolis, MN 55455, USA}
\email{plkelly@umn.edu}

\begin{abstract}

   We present 21 transients from our three-epoch, four-band NIRCam 
observations covering 14.16 arcmin$^2$ in the Spitzer IRAC Dark Field (IDF), 
taken by the JWST Prime Extragalactic Areas for Reionization and Lensing 
Science program with a time cadence of $\sim$6 months. A separate Hubble Space
Telescope program provided Advanced Camera for Surveys optical imaging 
contemporaneous with the second and third epochs of the NIRCam observations. The 
NIRSpec spectroscopy on three transients confirmed a Type Ia supernova at 
$z=1.63$ and the host galaxies of the other two at $z=2.64$ and 1.90, 
respectively. Combining these with the photometric redshifts ($z_{\rm ph}$) of 
the host galaxies in the rest of the sample, we find that the transients are 
in either a ``mid-$z$'' group at $z>1.6$ with $M_V\lesssim -16.0$ mag or a 
``low-$z$'' group at $z<0.4$ with 
$M_H\gtrsim -14.0$ mag. The mid-$z$ transients are consistent with supernovae.
In contrast, the low-$z$ transients’ luminosities fall in the range of the 
so-called ``gap transients'' between supernovae and novae. 
However, this latter conclusion is only tentative due to possible
catastrophic failures in $z_{\rm ph}$ that could bias them to low-$z$.
Conversely, if they are indeed at $z<0.4$, it would be worth studying similar
transients in the future. Our work further demonstrates the power of NIRCam in
transient science and also shows that it would be more fruitful to carry out
a long-term monitoring program with more passbands, a higher cadence and prompt
follw-up spectroscopy. Being in the continuous viewing zone of the JWST, the
IDF is an ideal field for this purpose.

\end{abstract}


\section{Introduction}

   Time-domain observations open new doorways to revealing the physical 
processes powering energetic astronomical phenomena. Such observations have 
been critical to our understanding of transient objects such as supernovae 
(SNe), gamma-ray bursts, tidal disruption events (TDEs), fast radio bursts, 
etc., as well as nontransient objects such as variable stars and active 
galactic nuclei (AGNs).

  The James Webb Space Telescope (JWST) is now recognized as an extremely
efficient facility for exploring the time-domain infrared sky. By comparing
the preimaging data for the NIRISS slitless observations of the GLASS-JWST
program \citep[][]{Treu2022} on the lensing cluster Abell 2744 ($z=0.308$) and
the archival Hubble Space Telescope (HST) WFC3 data in the same field,
\citet[][]{Chen2022GLASS} discovered a blue supergiant star at $z=2.65$, which 
manifested itself as a caustic transient in a highly magnified arc. The more 
prolific transient discovery machine is the NIRCam instrument. Based on the 
three-epoch NIRCam observations from the Prime Extragalactic Areas for 
Reionization and Lensing Science program \citep[PEARLS;][]{Windhorst2023} and 
with the aid of one more independent epoch from the Canadian NIRISS Unbiased 
Cluster Survey \citep[][]{Willott2022}, \citet[][]{Yan2023d} discovered 14 
transients in the lensing cluster field MACS J0416.1$-$2403 ($z=0.397$), 12 of 
which are caustic transits of individual stars in highly amplified background 
arcs and 2 are likely SNe at $z=2.205$ and 0.7093, respectively. By comparing 
the PEARLS NIRCam images on the $z=0.35$ lensing cluster PLCK~G165.7$+$67.0 
with the archival HST Advanced Camera for Surveys (ACS) data, 
\citet[][]{Frye2024} discovered the triply-imaged Type Ia supernova (SN~Ia) 
``SN~H0pe'', which was confirmed at $z=1.78$.

    Gravitational lensing is not a necessary condition for NIRCam's high
efficiency in transient discovery. \citet[][]{DeCoursey2025a} reported 79 SNe
in the GOODS-S field covering $\sim$25~arcmin$^2$, which were discovered as
transients mainly based on two epochs of NIRCam observations from the JWST
Advanced Deep Extragalactic Survey \citep[JADES;][]{Eisenstein2025} that
spanned a year. In terms of the highest redshifts probed, their sample
includes an SN~Ia at $z=2.90$ \citep[][]{Pierel2024c}, an SN~Ibc-BL at
$z=2.83$ \citep[][]{Siebert2024} and a likely SN~IIP at $z=3.613$ 
\citep[][]{Coulter2025}, all confirmed directly on the transients by follow-up 
spectroscopy. There are also a few transients associated with host galaxies 
confirmed at $z>4$, including a potential SN at $z=4.504$. Recently, 
\citet[][]{DeCoursey2025b} reported a possible SN (or a variable source) at the 
center of a $z=5.274$ galaxy in the GOODS-N field, found by comparing the 
three-band NIRCam data obtained by the Complete NIRCam Grism Redshift Survey 
program (PID 3577; PI: E. Egami) with those from the JADES program 1 yr 
earlier. Growing sample sizes of SNe at $z>2$ are opening up the possibility of 
studying the evolution of their properties 
\citep[e.g.,][]{Moriya2025, Pierel2025}, which will greatly impact their 
applications and the understanding of their host galaxies in general.
Moreover, SNe are not the only transient population being discovered by
NIRCam. For example, by comparing the four-band NIRCam images from the 
COSMOS-Web program \citep[][]{Casey2023} with the UltraVISTA archival near-IR 
data in the same area, \citet[][]{Karmen2025} discovered a candidate TDE at 
$z\approx 5$. All this highlights the great power of NIRCam as a transient 
discovery machine.

   The PEARLS program was the only one in the JWST Cycle~1 that had a 
built-in component by design to utilize NIRCam for time-domain science in 
``blank'' fields. For this purpose, it chose two areas in the JWST's 
continuous viewing zone around the north ecliptic pole (NEP), which can be 
visited by the JWST at any time. One area is the ``NEP Time-domain Field'' 
\citep[][]{Jansen2018}, for which PEARLS observed in eight NIRCam bands for one 
epoch to create the reference images for future repeated NIRCam observations by 
other programs. An HST transient and variability study has been carried out in 
this area as well \citep[][]{OBrien2024}. The other is the ``JWIDF'', which is 
in the central region of the historic Spitzer IRAC Dark Field
\citep[IDF;][]{Krick2009, Yan2018}. The IDF has the lowest zodiacal and
Galactic background in the sky (hence ``dark''); for this reason, it was the 
calibration field for the IRAC instrument on board the Spitzer Space Telescope 
and received nearly biweekly IRAC observations throughout the Spitzer mission
($\sim$16.2 yrs). PEARLS observed the JWIDF in four NIRCam bands in three 
epochs spanning the duration of Cycle 1. To facilitate transient studies in 
wavelength regimes not accessible by the JWST, we also obtained HST ACS 
observations contemporaneous with the last two NIRCam epochs. In total, we 
found 21 transients in 14.16~arcmin$^2$, three of which were observed 
spectroscopically by a follow-up NIRSpec program. Here, we present these 
results.

  Our paper is organized as follows. The NIRCam and ACS imaging observations
are described in Section~2. The transient search is presented in Section~3,
followed by a detailed description of the individual transients in Section~4. 
The NIRSpec spectroscopy of three transients is given in Section~5. Section~6 
discusses the results, and Section~7 gives a summary. All magnitudes are 
reported in the AB system unless noted otherwise. All coordinates are in the 
International Celestial Reference System frame (GAIA DR3, equinox J2000), and
distances where needed come from a flat $\Lambda$CDM cosmology with parameters
$H_0=71$~km~s$^{-1}$~Mpc$^{-1}$, $\Omega_{\rm M}=0.27$, and 
$\Omega_\Lambda=0.73$.

\section{Imaging Observations and Data Reduction}

\begin{table*}
\centering
\caption{Summary of the JWIDF NIRCam and contemporaneous ACS observations. }
\begin{tabular}{ccccc} \hline 
    Epoch & Instrument & Band:Pointing     & Obs.\ Start Time (UT)          &  Exp.\ time (s) \\
\hline 
    Ep1  &  NIRCam    & Four bands & 2022-07-08 03:23                & 3156.6 \\
\hline
    Ep2  &  NIRCam    & Four bands & 2023-01-06 08:54                & 2512.4 \\
        &  ACS  & F435W:E   & 2022-12-31 04:07  & 5205 \\
        &  ACS  & F435W:W   & 2023-01-05 01:27  & 5205 \\
        &       & F606W:E   & 2023-01-05 06:12  & 5100.7 \\
        &       & F606W:W   & 2023-01-05 23:39  & 5205 \\
        &       & F814W:E   & 2023-01-06 15:35  & 5205 \\
        &       & F814W:W   & 2023-01-07 02:36  & 5205 \\
\hline
    Ep3 &  NIRCam    & Four bands & 2023-07-06 22:03                & 2834.5 \\
        &  ACS  & F435W:W   & 2023-07-17 06:22 & 5205 \\
        &       & F435W:E   & 2023-07-25 05:49 & 6234 \\
        &       & F606W:W   & 2023-07-15 08:28 & 5205 \\
        &       & F606W:E   & 2023-07-19 15:33 & 5205 \\
        &       & F814W:E   & 2023-07-16 08:13 & 5205 \\ 
        &       & F814W:W   & 2023-07-16 11:22 & 5205 \\ 
        \hline 
    \end{tabular}
    \label{tab:obsinfo}
    \raggedright
    \tablecomments{(1) The total exposure times are given for the individual 
    pointings, which are the same in most cases but with two exceptions (Ep2 
    ACS F606W and Ep3 ACS F435W). (2) One visit of the Ep3 ACS F435W 
    observations failed on 2023 July 17 and was retaken on 2023 July 25. }
\end{table*}

\subsection{Three-epoch JWST/NIRCam Data}

    The JWIDF NIRCam data were taken in four bands, F150W and F200W in the 
short wavelength (SW) channel and F356W and F444W in the long wavelength (LW) 
channel. The instrument setups were detailed by \citet[][]{Yan2023b}. The 
resultant field is a contiguous $\sim$5\farcm9$\times$2\farcm4 rectangle with 
the gaps between detectors covered by dithered exposures.

    By design, the three-epoch observations spanned the JWST Cycle 1 duration, 
as summarized in Table~\ref{tab:obsinfo}. The first epoch (``Ep1'') was done at 
the beginning of the cycle, the second epoch (``Ep2'') was 6 months later, and
the third (``Ep3'') was after another 6 months. Because JWST's orientation 
rotates on a yearly basis, the two NIRCam modules were flipped 180\degr\ in Ep2 
as compared to Ep1 and Ep3. The time interval between Ep1 and Ep2 was 182.2 
days, while that between Ep2 and Ep3 was 181.5 days.

    The data reduction used the standard JWST science calibration pipeline
\citep[version 1.11.4, ][]{Bushouse24_jwppl} in the calibration context of 
{\tt jwst\_1130.pmap} starting from the Level 1b products retrieved from the 
Mikulski Archive for Space Telescopes (MAST)\null. Further steps followed the 
procedures  described by \citet[][]{Yan2023e}. The 180$^{\circ}$ flip of the 
NIRCam modules in Ep2 revealed small systematic flux calibration offsets among
the detectors, as reported by \citet[][]{Ma2024zp}
\footnote{These calibration offsets among the detectors are still present in
the latest {\tt jwst\_1364.pmap} as of this writing.}.
Because our transient
search relies on difference images between epochs (Section~3.1), such offsets 
must be corrected. Therefore, we used the recipe provided by 
\citet[][]{Ma2024zp} to remove these detector-dependent calibration offsets
before the final drizzling process. The astrometry was tied to the GAIA DR3,
with the help of the deep optical images in this field taken by the One Degree
Imager at the WYIN 3.6~m telescope (H. Yan et al., in preparation).
For this work, we produced the mosaics at the pixel scale of 0\farcs06. 
The corresponding AB magnitude zero-point is 26.581 when converting from the
surface brightness units of MJy~sr$^{-1}$ used by the JWST pipeline for all 
imaging data. We also derived the rms maps from the weight images using
{\sc astroRMS}
\footnote{Courtesy of M. Mechtley;  \url{https://github.com/mmechtley/astroRMS}},
which accounts for the correlated noise due to pixel resampling.

\subsection{Archival HST/ACS Data}

   A large portion of the IDF has archival HST/ACS data in F814W at a two-orbit 
depth \citep[][]{Krick2009} taken in 2006 (Cycle~14, PID 10521; PI J.\ Surace). 
These data cover the JWIDF and are referred to as ``Ep0.'' They were reduced 
together with the new ACS data as described below.

\subsection{Two-epoch Contemporaneous HST/ACS Data}
    
    The new HST ACS WFC data were taken by the HST Cycle 30 program ID~17154 
(PI H.\ Yan) in 24 orbits. The timing was specifically designed to be
contemporaneous (within $\pm$15 days, Table~\ref{tab:obsinfo}) with NIRCam Ep2 
and Ep3. We refer to these two ACS epochs as Ep2 and Ep3, respectively, 
despite there being no ACS observations contemporaneous with the NIRCam 
Ep1. In each epoch, the JWIDF NIRCam footprint was covered by two ACS 
pointings (labeled E and W) in the filters F435W, F606W, and F814W. Each epoch 
comprised two orbits of observation (split into six dithered exposures) at each 
pointing in each filter. We used a small dithering of 0\farcs033 to improve the
sampling of the ACS, at the expense of leaving unfilled gaps between the two CCD
chips.

    One of the visits for F435W in Ep3 was impacted by a camera anomaly and 
returned zero values in the data. This visit was reobserved as quickly as it 
could be scheduled, about 3 days beyond the planned time window, but these 
are still effectively contemporaneous.

\begin{figure*}
    \centering
    \includegraphics[width=\textwidth,height=\textheight,keepaspectratio]{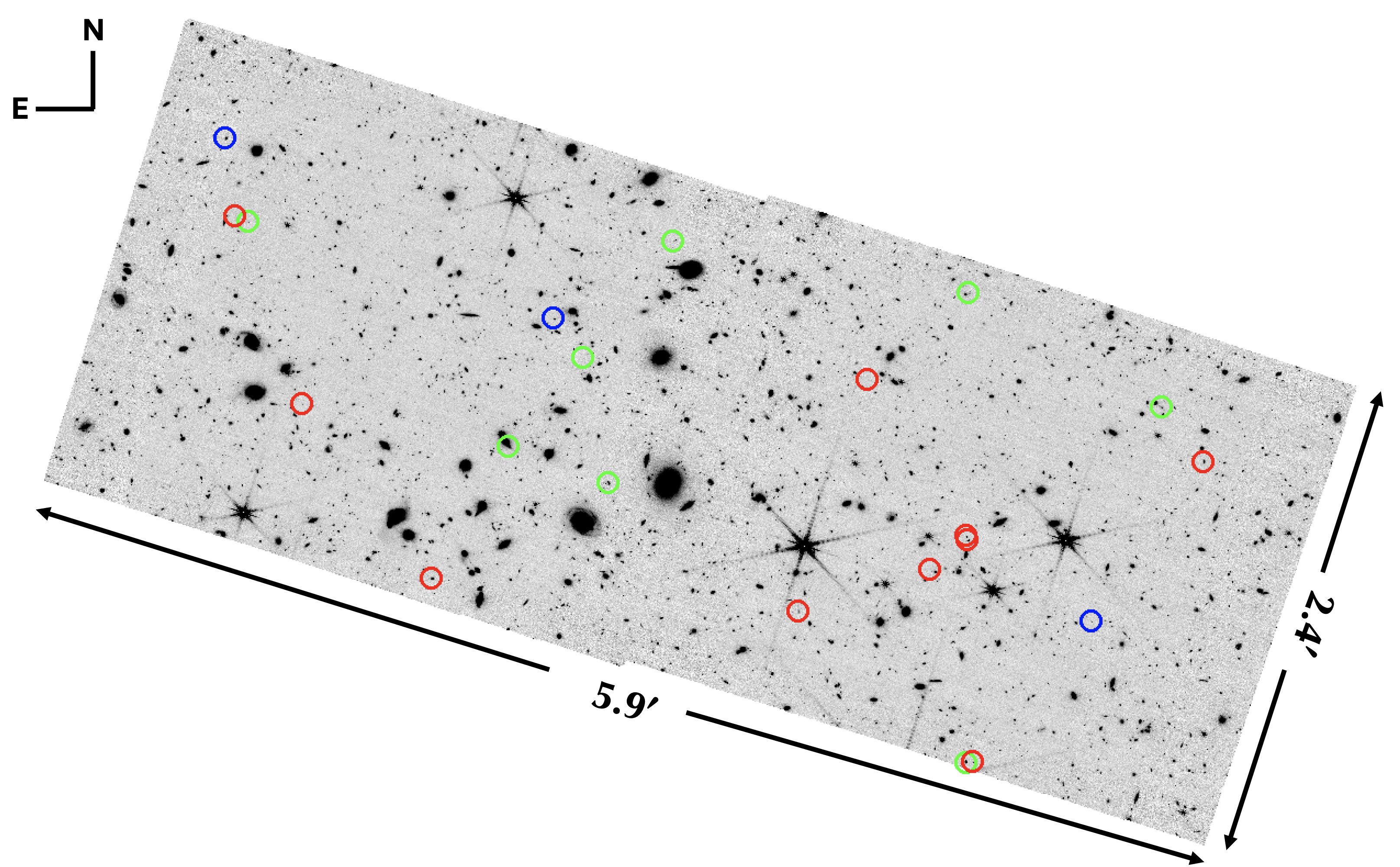}
    \raggedright
    \caption{Combined Ep1+Ep2+Ep3 F356W negative image of the JWIDF.  
    Image scale and orientation are indicated. The
    locations of the D13, D31, and D21 transients are marked by the red, 
    green, and blue circles, respectively.
    }
    \label{fig:transpos}
\end{figure*}

    To reduce the ACS data, we first processed the raw images using  
{\tt calacs} version 2.7.4 in the calibration context of {\tt hst\_1082.pmap}.
This step included subtraction of bias and dark current, correction for the 
deteriorated charge transfer efficiency, and flat-fielding. A known anomaly in 
the ACS WFC data is that the four quadrants can have slight background offsets.\footnote{\url{https://www.stsci.edu/hst/instrumentation/acs/performance/anomalies-and-artifacts}} 
As a remedy, we subtracted the median background value from each quadrant. To 
estimate the original per-quadrant background, we segmented each quadrant into 
sections of 128$\times$128 pixels, masked the sources, and applied a 3$\times$3 
median filter to obtain the median value. After the subtraction, all quadrants 
had zero background. We then registered single images to the same grid as 
defined by the NIRCam images and drizzled all the exposures in the same band 
using {\tt DrizzlePac} version 3.5.1. The magnitude zero-points for F435W, 
F606W, and F814W images are 25.6473, 26.4845 and 25.9390, respectively. The rms 
maps were also derived in the same way as in Section 2.1.
  
\section{Transients and Host Galaxies}
  
\subsection{Transient Search}

    Following the methodology of \citet[][]{Yan2023d}, we searched for
transients by detecting positive peaks on the F356W difference images between
the NIRCam epochs. The F356W band was chosen as the basis because (1) it is 
the most sensitive band in the LW channel, and (2) the LW channel is 
cosmetically much cleaner (e.g., free of ``snowballs,''
\footnote{\url{https://www.stsci.edu/files/live/sites/www/files/home/jwst/documentation/technical-documents/_documents/JWST-STScI-008545.pdf}},
``wisps,''
\footnote{\url{https://jwst-docs.stsci.edu/known-issues-with-jwst-data/nircam-known-issues/nircam-scattered-light-artifacts}},
etc.) than the SW band.

    The difference images were obtained in all bands by direct subtraction.
The corresponding RMS maps were also constructed by adding the RMS maps of the
involved bands in quadrature. The sets used to initiate the 
search (hereafter the ``searching sets'') were Ep1~$-$~Ep3 (hereafter
``D13'') for the transients that were decaying, Ep3~$-$~Ep1 (``D31'') for 
those that were emerging, Ep2~$-$~Ep1 (``D21'') and Ep2~$-$~Ep3 (``D23'') for
those that appeared in Ep2 but then became too faint or vanished in Ep3. 
Other combinations, such as Ep1~$-$~Ep2 (``D12''), Ep3~$-$~Ep2 (``D32''), etc., 
were only used for purposes such as constructing the spectral energy 
distributions (SEDs) once the transients were located.

    For each of the four search images, we ran {\sc SExtractor} 
\citep[][]{Bertin1996} in dual-image mode with F356W as the detection band. 
A 5$\times5$ pixel Gaussian filter with a 2 pixel full width at 
half-maximum was used to convolve the detection image, the threshold was set
to 1.0~$\sigma$, and at least 4 connected pixels above this threshold are
required. We adopted the \texttt{MAG\_ISO} magnitudes, and the magnitude 
uncertainties were calculated using the associated RMS maps. Only the peaks 
having signal-to-noise ration ($\rm{S/N}$) $\geq 5$ in F356W were considered, 
giving $\sim$900 sources in each searching set. For reference, the 2$\sigma$ 
limits (measured in a circular aperture with radius $r=0\farcs2$) of the D13 
and D31 sets are 28.42, 28.62, 29.14, 28.87~mag in F150W, F200W, F356W, and 
F444W, respectively, and those of the D21 set are 28.32, 28.51, 29.05, and 
28.77~mag, respectively.

    After the initial search, all transient candidates were visually examined 
to reject contaminants. Most of those were artifacts caused by imperfect 
subtraction of bright stars and galaxies. In the end, 21 unique transients 
survived: 10 from D13, 8 from D31, and the same 3 from both D21 and D23.
All 21 are pointlike, meaning their angular diameters are $<$0\farcs1. 
As discussed below, all three transients found D21/D23 transients were still 
visible in the Ep3 images, i.e., they decayed by Ep3 but did not completely 
disappear. (We will refer to these three as the ``D21 transients.'')
Figure~\ref{fig:transpos} shows the spatial distribution of the 21 transients.

    5 of the 21 transients are ``hostless'' (3 in D13, 2 in D31, and 
none in D21), that is, they have no host galaxies visible in the NIRCam images. 
The other 16 split evenly: 8 of them are deeply embedded in their host 
galaxies and would not be detectable in the original images, and 8 are at 
the outskirts of their hosts and could be discerned in the original images. 
We refer to specific transients as ``T-D??-x'', where ``D??'' is the discovery
set (i.e., ``D13'', ``D31'' and ``D21'') and ``x'' is a letter running 
alphabetically starting from ``A''.

\subsection{Transient Photometry and SEDs}
\label{s:tphot}

   For the five hostless transients, we ran {\sc SExtractor} in dual-image mode 
on the original images to obtain matched-aperture photometry for all bands 
(NIRCam and ACS) in each epoch. The discovery F356W difference images were used 
as the detection images, and the \texttt{MAG\_ISO} magnitudes were adopted to 
construct their per-epoch SEDs. Because there no Ep1 ACS data, the SEDs of the 
D13 hostless transients (three of them) in Ep1 do not extend to the ACS bands 
(but their Ep2 and Ep3 SEDs do). The situation for the D31 hostless transients 
(two of them) is similar, but the archival F814W images Ep0 images give upper 
limits in that band. 

   For the 16 transients having visible host galaxies, photometry had to be 
done on difference images, where the host contamination was removed. This 
approach assumes that each transient was invisible in the reference image so 
that its brightness is not erroneously suppressed by the subtraction. To study 
how the sources' SEDs changed over time, the photometry was done on sets of 
difference images that could track the full time evolution. Specifically, 
(1) for the D13 transients, the D13 set and the D23 set gave SEDs in Ep1 and 
Ep2, respectively, assuming that the transients were invisible in the Ep3 
images; (2) for the D31 transients, the D21 and D31 sets gave SEDs in Ep2 and 
Ep3, respectively, assuming that the transients were invisible in Ep1; (3) for 
the D21 transients, the D21 set and the D31 set were used to obtain their SEDs 
in Ep2 and Ep3, respectively, assuming that they are invisible in Ep1. Again, 
the ACS data (contemporaneous and archival) were also used when available. 

\begin{figure*}
    \centering
    \includegraphics[width=0.9\textwidth]{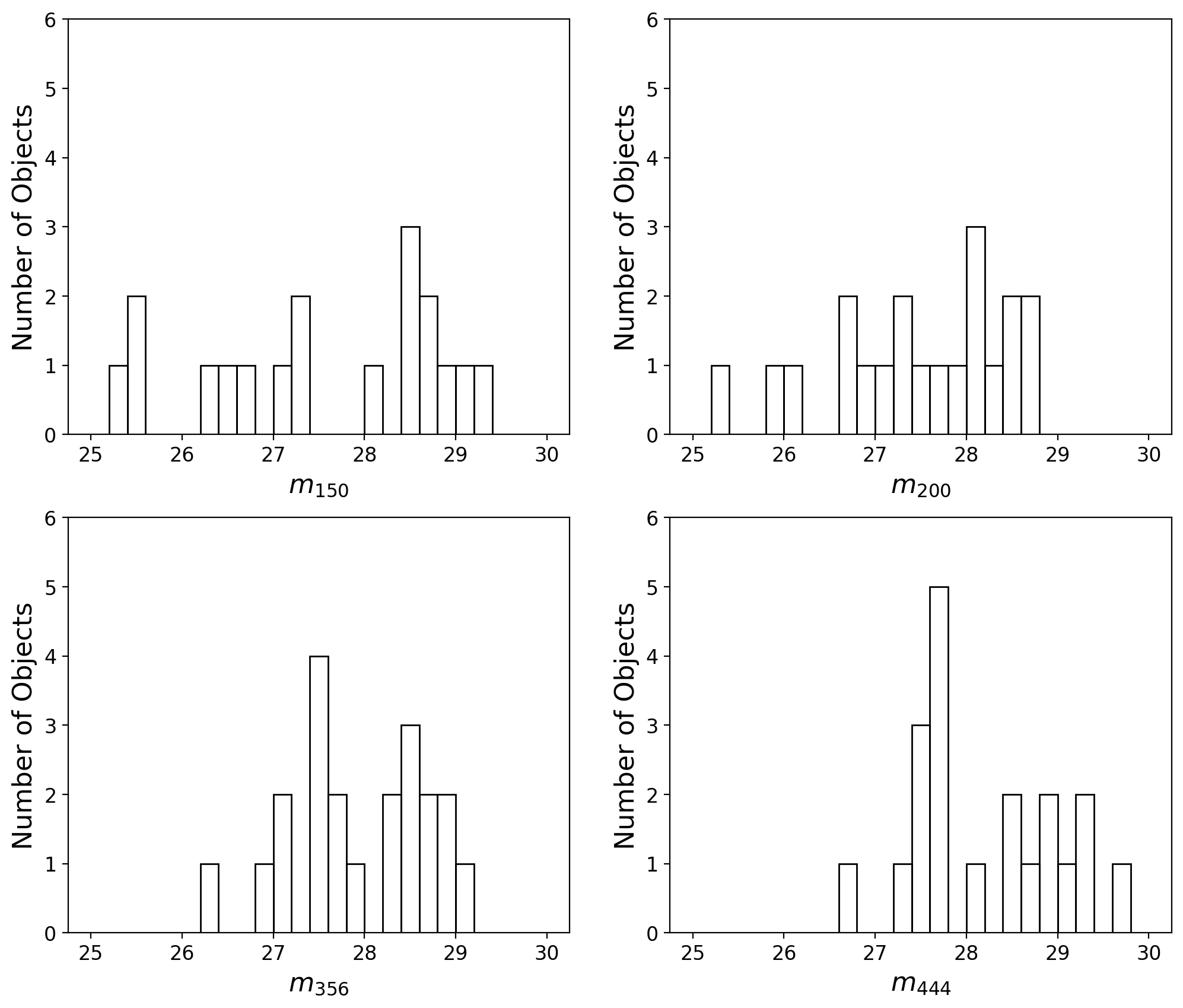}
    \raggedright
    \caption{Distribution of the NIRCam magnitudes for all transients in their discovery epoch. Magnitude limits are not included. (By construction there are no limits in the $m_{356}$ panel.)}
    \label{fig:maghist}
\end{figure*}

    All the photometric information is reported in Tables~\ref{tab:catD13},
\ref{tab:catD31}, and \ref{tab:catD21}. Figure~\ref{fig:maghist} shows the
magnitude distributions in various bands. Because of the sparse, 6 month 
cadence, most of our 21 transients have high-S/N SEDs only in the discovery 
epoch. Interestingly, their SEDs can be well described by piecewise power law, 
which is shown in Appendix~A.

\begin{table*}[ht]
\tiny
\centering
\caption{Catalog of the JWIDF D13 transients}
\label{tab:catD13}
\begin{tabular}{ccccccccccc}
\hline
ID & R.A. (deg) & Decl. (deg) & $m_{435}$ & $m_{606}$ & $m_{814}$ & $m_{150}$ & $m_{200}$ & $m_{356}$ & $m_{444}$ & Epoch \\
\hline
T-D13-D & 264.9836098 & 68.9629839 & \ldots & \ldots & \ldots & $29.02\pm0.20$ & $28.79\pm0.14$ & $29.02\pm0.12$ & $29.68\pm0.27$ & Ep1 \\
        &             &            & ${>}28.10$ & ${>}28.74$ & ${>}28.23$ & ${>}29.10$ & $29.53\pm0.32$ & $30.45\pm0.52$ & ${>}29.41$ & Ep2 \\
        &             &            & ${>}28.05$ & ${>}28.52$ & ${>}28.01$ & $29.48\pm0.33$ & $29.73\pm0.35$ & ${>}29.78$ & $30.12\pm0.47$ & Ep3 \\
T-D13-K & 264.9975851 & 68.9782033 & \ldots & \ldots & \ldots & $28.47\pm0.16$ & $28.16\pm0.09$ & $28.39\pm0.09$ & $28.51\pm0.11$ & Ep1 \\
        &             &            & ${>}28.11$ & ${>}28.72$ & $28.43\pm0.35$ & ${>}28.62$ & ${>}28.83$ & $30.04\pm0.46$ & ${>}29.09$ & Ep2 \\
        &             &            & ${>}27.97$ & ${>}28.53$ & ${>}27.98$ & ${>}29.05$ & ${>}29.30$ & $30.22\pm0.49$ & ${>}29.30$ & Ep3 \\
T-D13-L & 265.1388289 & 68.9913262 & \ldots & \ldots & \ldots & $27.13\pm0.07$ & $27.17\pm0.06$ & $27.65\pm0.06$ & $27.69\pm0.09$ & Ep1 \\
        &             &            & ---      & ---      & ---      & ${>}28.59$ & ${>}28.78$ & $29.49\pm0.48$ & ${>}29.03$ & Ep2 \\
        &             &            & ${>}27.73$ & ${>}28.54$ & ${>}28.00$ & ${>}28.97$ & ${>}29.18$ & ${>}29.70$ & ${>}29.37$ & Ep3 \\
\hline
T-D13-C & 265.0949466 & 68.9622828 & \ldots & \ldots & \ldots & $26.21\pm0.07$ & $26.76\pm0.09$ & $27.66\pm0.10$ & $27.52\pm0.12$ & Ep1 \\
        &             &            & ${>}27.57$ & ${>}28.36$ & ${>}27.86$ & $27.98\pm0.34$ & ${>}28.41$ & ${>}29.07$ & ${>}28.76$ & Ep2 \\
T-D13-E & 264.9752185 & 68.9654010 & \ldots & \ldots & \ldots & $27.22\pm0.09$ & $26.97\pm0.07$ & $28.94\pm0.20$ & ${>}29.13$ & Ep1 \\
        &             &            &  ---  & ---  & --- & ${>}28.37$ & ${>}28.53$ & ${>}29.36$ & ${>}29.02$ & Ep2 \\
T-D13-I & 264.9224693 & 68.9715861 & \ldots & \ldots & \ldots & $27.33\pm0.11$ & $27.44\pm0.10$ & $28.38\pm0.15$ & $28.97\pm0.34$ & Ep1 \\
        &             &            & ${>}27.79$ & ${>}28.36$ & ${>}27.82$ & $28.02\pm0.25$ & $27.59\pm0.14$ & $28.44\pm0.17$ & $28.80\pm0.31$ & Ep2 \\
\hline
T-D13-A & 264.9740158 & 68.9476210 & \ldots & \ldots & \ldots & $28.83\pm0.44$ & ${>}28.55$ & $28.50\pm0.16$ & $28.50\pm0.22$ & Ep1 \\
        &             &            & ${>}27.77$ & ${>}28.37$ & ${>}27.82$ & ${>}28.31$ & ${>}28.50$ & ${>}29.04$ & ${>}28.73$ & Ep2 \\
T-D13-B & 265.0130451 & 68.9597128 & \ldots & \ldots & \ldots & $28.62\pm0.28$ & $28.45\pm0.20$ & $28.66\pm0.17$ & $29.16\pm0.34$ & Ep1 \\
        &             &            & ${>}27.75$ & ${>}28.32$ & ${>}27.77$ & ${>}28.57$ & ${>}28.76$ & $29.14\pm0.30$ & $29.10\pm0.37$ & Ep2 \\
T-D13-F & 264.9754899 & 68.9657026 & \ldots & \ldots & \ldots & $26.75\pm0.09$ & $27.20\pm0.12$ & $27.51\pm0.08$ & $27.62\pm0.11$ & Ep1 \\
        &             &            & --- & --- & --- & ${>}28.36$ & ${>}28.51$ & $28.92\pm0.31$ & ${>}29.06$ & Ep2 \\
T-D13-J & 265.1238690 & 68.9762653 & \ldots & \ldots & \ldots & $28.49\pm0.39$ & $28.43\pm0.31$ & $28.51\pm0.14$ & $28.98\pm0.28$ & Ep1 \\
        &             &            & ${>}27.54$ & ${>}28.32$ & ${>}27.77$ & ${>}28.31$ & ${>}28.55$ & ${>}29.29$ & ${>}29.03$ & Ep2 \\
\hline
\end{tabular}
\raggedright
\tablecomments{(1) The transients are arranged in the order of their 
categories (``hostless,'' ``outskirt,'' and ``deeply embedded''), separated by
horizontal lines. 
(2) For a nondetection (${\rm S/N<2}$), the 2$\sigma$ upper limit measured 
within a circular aperture of 0\farcs2 in radius is quoted. 
(3) For the hostless transients, the photometry was done on the original 
images and therefore is available for three epochs. 
(4) For the other two categories, the photometry was done on the difference 
images using the Ep3 images as the reference and therefore is only available 
for Ep1 and Ep2 (D13 for Ep1 and D23 for Ep2, respectively). 
(5) In Ep1, the photometry did not extend to the ACS bands because the ACS 
data were not taken in that epoch.
(6) \texttt{T-D13-L} has no ACS photometry in Ep2 because it fell in the gap of
the ACS images in this epoch.
(7) \texttt{T-D13-E} and \texttt{T-D13-F} are very close to each other; they 
have no ACS photometry because they fell in the gap of the Ep3 ACS images, and
therefore, no difference ACS images could be constructed.
}
\end{table*}

\begin{table*}[hbt!]
\tiny
\centering
\caption{Catalog of the JWIDF D31 transients }
\label{tab:catD31}
\begin{tabular}{ccccccccccc}
\hline
ID & RA (deg) & Dec (deg) & $m_{435}$ & $m_{606}$ & $m_{814}$ & $m_{150}$ & $m_{200}$ & $m_{356}$ & $m_{444}$ & Epoch \\
\hline
T-D31-E & 265.0609529 & 68.9799794 & \ldots & \ldots & \ldots & ${>}29.13$ & ${>}29.29$ & ${>}29.72$ & ${>}29.46$ & Ep1 \\
        &             &            & ${>}28.09$ & ${>}28.73$ & ${>}28.22$ & ${>}28.88$ & ${>}29.04$ & ${>}29.50$ & ${>}29.20$ & Ep2 \\
        &             &            & ${>}27.68$ & $28.82\pm0.53$ & $25.96\pm0.06$ & $25.44\pm0.02$ & $25.98\pm0.02$ & $27.05\pm0.04$ & $27.41\pm0.08$ & Ep3 \\

T-D31-F & 264.9750417 & 68.9851622 & \ldots & \ldots & \ldots & $29.10\pm0.35$ & $28.63\pm0.17$ & $29.40\pm0.30$ & $29.35\pm0.33$ & Ep1 \\
        &             &            & ${>}28.17$ & ${>}28.76$ & ${>}28.33$ & ${>}28.20$ & $28.60\pm0.39$ & $28.90\pm0.22$ & ${>}29.04$ & Ep2 \\
        &             &            & ${>}28.07$ & ${>}28.61$ & ${>}28.07$ & $26.41\pm0.03$ & $26.69\pm0.03$ & $27.88\pm0.07$ & $27.75\pm0.08$ & Ep3 \\
\hline

T-D31-A & 264.9756537 & 68.9475589 & \ldots & \ldots & ${>}28.10$ & ${>}28.34$ & ${>}28.52$ & ${>}29.00$ & ${>}28.69$ & Ep2 \\
        &             &            & ${>}27.77$ & ${>}28.34$ & ${>}27.97$ & $28.43\pm0.30$ & $28.11\pm0.19$ & $28.43\pm0.16$ & $28.71\pm0.28$ & Ep3 \\

T-D31-C & 265.0777675 & 68.9729003 & \ldots & \ldots & ${>}28.16$ & ${>}28.71$ & ${>}28.95$ & ${>}29.44$ & $28.98\pm0.39$ & Ep2 \\
        &             &            & ${>}27.52$ & ${>}28.27$ & ${>}27.96$ & $28.05\pm0.26$ & $27.60\pm0.14$ & $27.53\pm0.07$ & $27.76\pm0.12$ & Ep3 \\

T-D31-G & 265.0409558 & 68.9893575 & \ldots & \ldots & ${>}28.12$ & ${>}28.62$ & ${>}28.79$ & ${>}29.01$ & ${>}28.64$ & Ep2 \\
        &             &            & ${>}27.56$ & ${>}28.31$ & ${>}28.00$ & ${>}28.57$ & $27.97\pm0.18$ & $27.59\pm0.09$ & $28.08\pm0.19$ & Ep3 \\
\hline

T-D31-B & 265.0553171 & 68.9699472 & \ldots & \ldots & ${>}28.17$ & ${>}28.06$ & ${>}28.30$ & ${>}28.85$ & ${>}28.55$ & Ep2 \\
        &             &            & $27.60\pm0.54$ & $26.99\pm0.15$ & $25.65\pm0.06$ & $25.26\pm0.04$ & $25.34\pm0.04$ & $26.33\pm0.05$ & $26.70\pm0.10$ & Ep3 \\

T-D31-D & 264.9317621 & 68.9760033 & \ldots & \ldots & ${>}28.21$ & ${>}28.60$ & ${>}28.79$ & ${>}29.36$ & ${>}29.10$ & Ep2 \\
        &             &            & ${>}27.73$ & ${>}28.34$ & ${>}27.97$ & $29.21\pm0.45$ & $28.70\pm0.23$ & $28.84\pm0.17$ & $29.36\pm0.38$ & Ep3 \\

T-D31-H & 265.1359783 & 68.9909039 & --- & --- & --- & $28.36\pm0.32$ & $28.19\pm0.22$ & $28.75\pm0.22$ & $28.97\pm0.35$ & Ep2 \\
        &             &            & \ldots & \ldots & ${>}27.88$ & $28.79\pm0.39$ & $28.23\pm0.19$ & $28.63\pm0.17$ & $29.36\pm0.44$ & Ep3 \\
\hline
\end{tabular}
\raggedright
\tablecomments{Similar to Table~\ref{tab:catD13} but for the D31 transients. 
(1) The photometry for the two hostless transients was done on the original 
images in the three epochs and therefore is available for all the three epochs 
(but not in the Ep1 ACS bands due to the lack of ACS data in Ep1). 
(2) For the outskirt and deeply embedded transients, their NIRCam photometry 
was done on the difference images using the Ep1 images as the reference and 
therefore is only available for Ep2 and Ep3 (D21 for Ep2 and D31 for Ep3, 
respectively). Their ACS F814W photometry was done on the difference images
using the PID 10521 (``Ep0'') F814W image as the reference and therefore is 
available for both Ep2 and Ep3 in most cases. Their ACS F435W and F606W 
photometry is only available in Ep3 because the Ep2 ACS images had to be used as 
the references in constructing the D32 difference images. 
(3) \texttt{T-D31-H} (in the deeply embedded category) has no photometry in 
Ep2 F814W because this transient fell in the gap of the ACS image. It has no
photometry in Ep3 F435W and F606W for the same reason (lacking the references
to construct the D32 difference images).
}
\end{table*}

\begin{table*}[hbt!]
\tiny
\centering
\caption{Catalog for the D21 transients}
\label{tab:catD21}
\begin{tabular}{ccccccccccc}
\hline
ID & RA (deg) & Dec (deg) & $m_{435}$ & $m_{606}$ & $m_{814}$ & $m_{150}$ & $m_{200}$ & $m_{356}$ & $m_{444}$ & Epoch \\
\hline
T-D21-A & 264.9474962 & 68.9588681 & ${>}28.01$ & $26.77\pm0.08$ & $24.85\pm0.03$ & $25.42\pm0.02$ & $26.17\pm0.04$ & $26.98\pm0.05$ & $27.38\pm0.10$ & Ep2 \\
        &             &            & ${>}28.07$ & ${>}28.52$       & ${>}28.02$       & $28.18\pm0.31$ & ${>}28.84$       & $29.34\pm0.47$ & ${>}29.12$       & Ep3 \\
\hline
T-D21-B & 265.0675752 & 68.9831352 & ${>}28.08$ & ${>}28.68$       & ${>}28.24$       & ${>}28.83$    & $28.16\pm0.18$ & $27.54\pm0.07$ & $27.69\pm0.10$ & Ep2 \\
        &             &            & ${>}27.71$ & ${>}28.56$       & ${>}28.02$       & ${>}28.98$    & ${>}29.15$       & $28.42\pm0.14$ & $28.80\pm0.26$ & Ep3 \\
T-D21-C & 265.1410616 & 68.9975505 & ${>}27.51$ & ${>}28.35$       & ${>}28.15$       & ${>}27.96$    & $27.22\pm0.18$ & $27.18\pm0.10$ & $27.43\pm0.17$ & Ep2 \\
        &             &            & \ldots   & \ldots         & ${>}27.95$       & ${>}28.23$    & $27.82\pm0.24$ & ${>}28.92$       & $28.38\pm0.36$ & Ep3 \\
\hline
\end{tabular}
\raggedright
\tablecomments{Similar to Table~\ref{tab:catD13} but for the D21 transients.
One of them is in the ``outskirt'' category and the other two are in the
``deeply embedded'' category, separated by the horizontal line.
(1) The NIRCam photometry was done on the difference images using the Ep1 
images as the references and is available for Ep2 and Ep3.
(2) In most cases, the ACS F814W photometry was done on the difference images 
using the PID 10521 (``Ep0'') F814W image as the reference and is available for 
Ep2 and Ep3.
(3) For \texttt{T-D21-A}, its host is invisible in the ACS F435W and F606W, and
therefore, its photometry in these two bands were measured on the original Ep2
and Ep3 images directly.
(4) \texttt{T-D21-B} and its host are invisible in all ACS bands, and therefore,
all the ACS upper limits were derived using the original ACS images.
(5) For \texttt{T-D21-C}, we assumed that this transient disappeared from the
ACS in Ep3, and the Ep2 upper limits in F435W and F606W were derived using the
D23 difference images.
}
\end{table*}

\subsection{Photometry of Host Galaxies and SED Fitting}

   To facilitate the understanding of these transients, we also obtained 
photometry of the host galaxies of the 16 transients having hosts. In the 
NIRCam bands, this was done in the epoch in which the transient was assumed 
invisible (e.g., using the Ep1 images for the D31 transients). In the ACS F814W 
band, the host photometry was done using the archival Ep0 data. The treatment 
for the ACS F435W and F606W bands was more complicated. Two of the D13 
transients, \texttt{T-D13-E} and \texttt{T-D13-F}, which are close to each 
other, fell in the gap of the Ep3 ACS data, and we had to use the Ep2 ACS 
images for them. Based on the Ep2 NIRCam images, the transients should 
have decayed by then. For the hosts of the D31 transients, we also had to use 
the Ep2 images. As luck would have it, this left one D31 transient, 
\texttt{T-D31-H}, in the unfilled gap, for which we had to use the Ep3 ACS 
images. Fortunately, this particular transient was invisible in all ACS bands, 
and the host magnitudes should be accurate. For the hosts of the D21 
transients, we used the Ep3 ACS images. Based on the shape of the 
transients' SEDs, the transients were likely invisible in these images.

    For all the above, {\sc SExtractor} was run in the dual-image mode with the
F356W image in the proper epoch as the detection image, and the
\texttt{MAG\_ISO} magnitudes were adopted as usual. For any non-detections
(${\rm S/N}<2$), the upper limits measured within a 0\farcs2 radius aperture
on the RMS maps were adopted. The results are given in 
Table~\ref{tab:phot_hosts}.

\begin{figure*}[hbt!]
    \centering
    \includegraphics[width=\textwidth,height=\textheight,keepaspectratio]{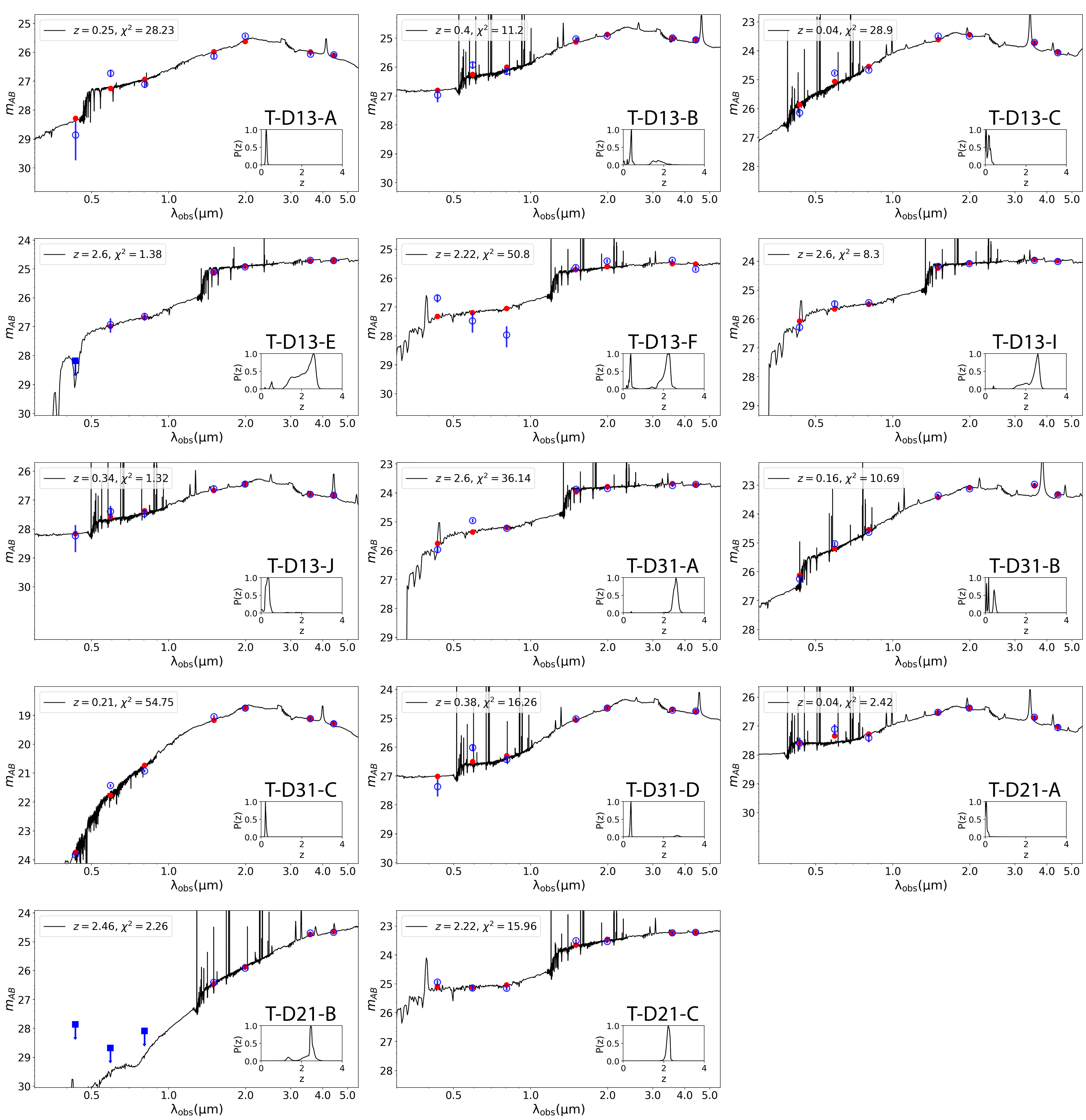}
    \raggedright
    \caption{\texttt{EAZY} SED-fitting results for the host galaxies of  
    transients that have detectable hosts. The IDs of the corresponding 
    transients are indicated. In each panel, the blue symbols (circles with 
    error bars and upper limits) show the observations, the curve is the 
    best-fit template, and the red symbols are the synthesized magnitudes based 
    on the best-fit template. The derived photometric redshift ($z_{\rm ph}$) 
    and the $\chi^2$ value of the fit are indicated in each panel, and the
    inset shows the probability distribution of $z_{\rm ph}$.  The hosts of 
    \texttt{T-D31-G} and \texttt{T-D31-H} are not included because they have 
    spectroscopic redshifts (Section 5.2). 
    \vspace{1cm}
    }
    \label{fig:sed_hosts}
\end{figure*}

    To obtain the photometric redshifts ($z_{\rm ph}$) of 
the host galaxies, we ran {\sc EAZY}-py
\citep[version 0.8.5;][]{Brammer2008} on the host galaxy SEDs. We used the 
``fsps\_QSF\_12\_v3'' templates, which are based on the flexible stellar 
population synthesis (FSPS) models of \citet{Conroy2010}. The redshift was 
allowed to vary between 0 and 10, and the results are summarized in
Figure~\ref{fig:sed_hosts}.  \texttt{T-D31-G} and \texttt{T-D31-H}
are not included because they have spectroscopic redshifts (Section 5.2).
A caveat in these $z_{\rm ph}$ estimates is that some of their redshift
probability distribution functions (PDF($z$)) are rather broad and/or have
multiple peaks. This is largely due to the limited number of passbands: we only
have four NIRCam bands and three ACS bands, and there is a wide gap of 
$\sim$0.7~$\mu$m between the bluest NIRCam band (F150W) and the reddest ACS
band (F814W). For this reason, these $z_{\rm ph}$ should be used with caution.

\begin{table*}
\centering
\tiny
\caption{Photometry of the transients' host galaxies}
\begin{tabular}{lccccccccc}
\hline
Transient ID & RA$_{\rm h}$ (deg) & Decl$_{\rm h}$ (deg) & $m_{\rm 435}$ & $m_{\rm 606}$ & $m_{\rm 814}$ & $m_{\rm 150}$ & $m_{\rm 200}$ & $m_{\rm 356}$ & $m_{\rm 444}$\\
\hline
T-D13-C & 265.094813 & 68.962335 & $26.14\pm0.15$ & $24.76\pm0.04$ & $24.66\pm0.03$ & $23.48\pm0.02$ & $23.48\pm0.01$ & $23.71\pm0.01$ & $24.06\pm0.02$ \\
T-D13-E & 264.975147 & 68.965364 & $>$28.03       & $26.92\pm0.24$ & $26.64\pm0.12$ & $25.11\pm0.04$ & $24.93\pm0.03$ & $24.70\pm0.01$ & $24.71\pm0.02$ \\
T-D13-I & 264.922505 & 68.971693 & $26.28\pm0.12$ & $25.47\pm0.08$ & $25.43\pm0.03$ & $24.18\pm0.02$ & $24.08\pm0.02$ & $23.97\pm0.01$ & $24.01\pm0.01$ \\
T-D13-A & 264.974050 & 68.947614 & $28.86\pm0.60$ & $26.72\pm0.11$ & $27.10\pm0.11$ & $26.13\pm0.07$ & $25.43\pm0.03$ & $26.06\pm0.03$ & $26.08\pm0.04$ \\
T-D13-B & 265.013057 & 68.959732 & $26.96\pm0.22$ & $25.92\pm0.11$ & $26.15\pm0.09$ & $25.02\pm0.04$ & $24.92\pm0.03$ & $24.99\pm0.02$ & $25.05\pm0.03$ \\
T-D13-F & 264.975487 & 68.965676 & $26.68\pm0.13$ & $27.48\pm0.33$ & $27.96\pm0.35$ & $25.63\pm0.06$ & $25.40\pm0.04$ & $25.38\pm0.02$ & $25.69\pm0.03$ \\
T-D13-J & 265.123870 & 68.976277 & $28.23\pm0.44$ & $27.39\pm0.21$ & $27.46\pm0.14$ & $26.60\pm0.10$ & $26.44\pm0.07$ & $26.80\pm0.04$ & $26.83\pm0.06$ \\ 
\hline 
T-D31-A & 264.975777 & 68.947651 &
$25.96\pm0.08$ & $24.95\pm0.04$ & $25.22\pm0.04$ &
$23.87\pm0.02$ & $23.85\pm0.01$ & $23.74\pm0.01$ & $23.70\pm0.01$ \\
T-D31-C & 265.078607 & 68.973360 &
$23.82\pm0.05$ & $21.43\pm0.01$ & $20.92\pm0.01$ &
$19.04\pm0.01$ & $18.74\pm0.01$ & $19.11\pm0.01$ & $19.29\pm0.01$ \\
T-D31-G & 265.040574 & 68.989442 &
$>$28.13 & $26.89\pm0.19$ & $27.08\pm0.11$ &
$25.93\pm0.06$ & $25.61\pm0.04$ & $25.55\pm0.02$ & $25.53\pm0.03$ \\
T-D31-B & 265.055412 & 68.969994 &
$26.24\pm0.13$ & $25.02\pm0.05$ & $24.63\pm0.02$ &
$23.35\pm0.02$ & $23.12\pm0.01$ & $22.97\pm0.01$ & $23.33\pm0.01$ \\
T-D31-D & 264.931863 & 68.976018 &
$27.36\pm0.29$ & $26.01\pm0.10$ & $26.43\pm0.08$ &
$25.02\pm0.03$ & $24.64\pm0.02$ & $24.72\pm0.01$ & $24.75\pm0.02$ \\
T-D31-H & 265.135985 & 68.990871 &
$>$27.71 & $26.97\pm0.18$ & $26.84\pm0.13$ &
$26.08\pm0.05$ & $26.03\pm0.04$ & $26.18\pm0.03$ & $26.19\pm0.04$ \\
\hline
T-D21-A & 264.947628 & 68.958876 &
$27.59\pm0.20$ & $27.11\pm0.18$ & $27.40\pm0.14$ &
$26.52\pm0.07$ & $26.37\pm0.05$ & $26.69\pm0.04$ & $27.05\pm0.08$ \\
T-D21-B & 265.067566 & 68.983132 &
${>}27.71$ & ${>}$28.57 & ${>}28.03$ &
$26.40\pm0.07$ & $25.91\pm0.04$ & $24.70\pm0.01$ & $24.68\pm0.01$ \\
T-D21-C & 265.141018 & 68.997571 &
$24.94\pm0.05$ & $25.14\pm0.06$ & $25.15\pm0.04$ &
$23.50\pm0.02$ & $23.53\pm0.01$ & $23.23\pm0.01$ & $23.22\pm0.01$ \\
\hline
\end{tabular}
\raggedright
\tablecomments{(1) The five ``hostless'' transients are not included in this
table. (2) RA$_{\rm h}$ and Decl$_{\rm h}$ are the coordinates of the centroids 
of the host galaxies.}
\label{tab:phot_hosts}
\end{table*}

\section{Notes on Individual Transients}

\begin{figure*}
    \centering
    \includegraphics[height=0.90\textheight,keepaspectratio]{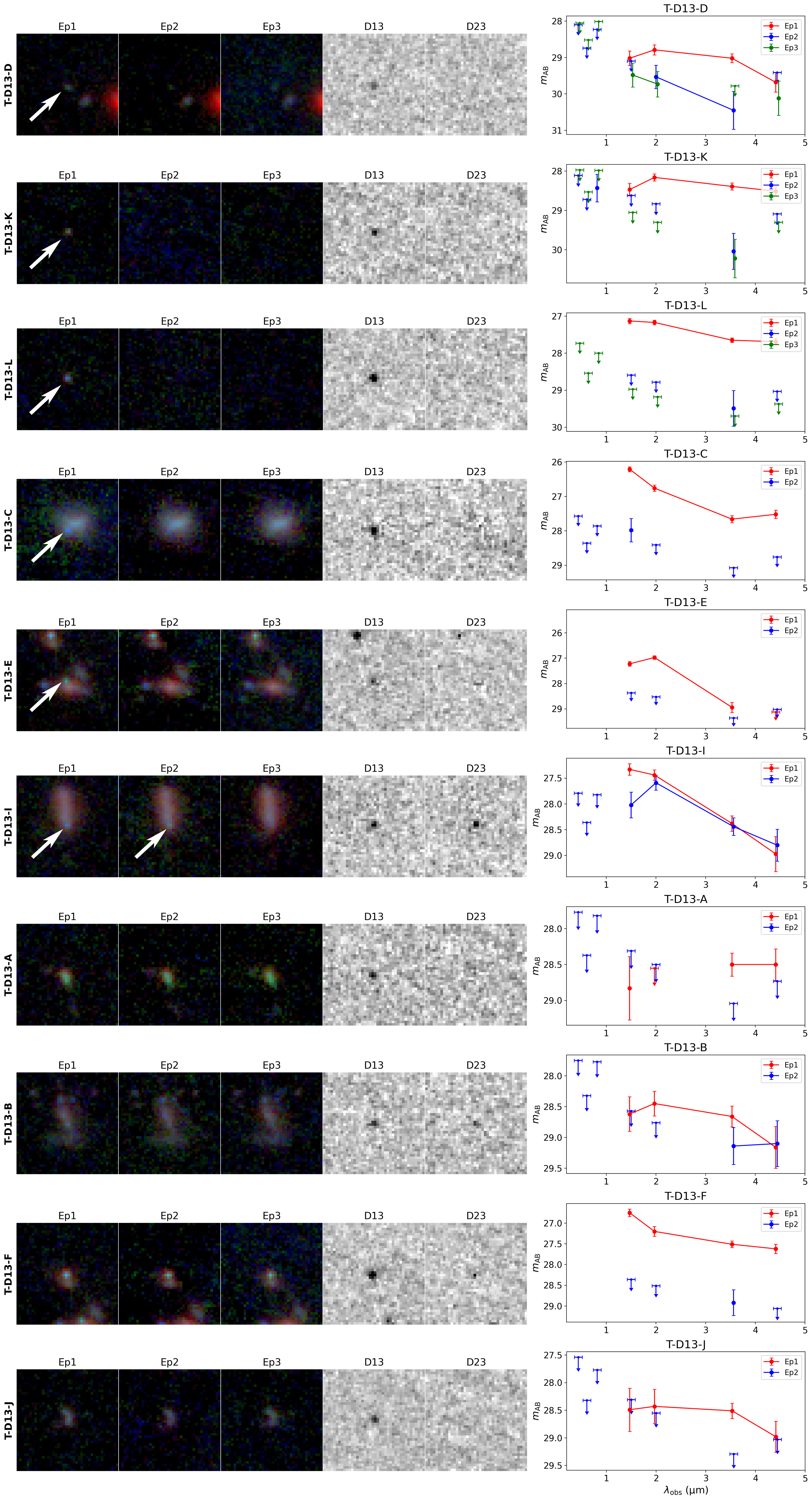}
    \caption{Image stamps and SEDs for D13 transients. All images are 2\farcs4 
    on a side and have north up, east to the left. For each object, the left 
    three panels show  NIRCam color image stamps in each epoch. These have F150W 
    as blue, F200W as green, and F356W + F444W as red. The next two panels show 
    negative D13 and D23 difference images. The right panel show the per-epoch 
    SEDs with points for different epochs shifted by $0.06~\mu$m  
    for clarity. Red circles show the Ep1 (discovery) SED,  blue triangles show 
    the Ep2 SED, and for the two hostless transients, green squares show the Ep3 
    SED. As mentioned in Section~\ref{s:tphot}, the photometry for transients 
    that have visible hosts assumes the transient had zero flux in Ep3.
    }
    \label{fig:t-d13-all}
\end{figure*}

\begin{figure*}
    \centering
    \includegraphics[width=\textwidth,height=\textheight,keepaspectratio]{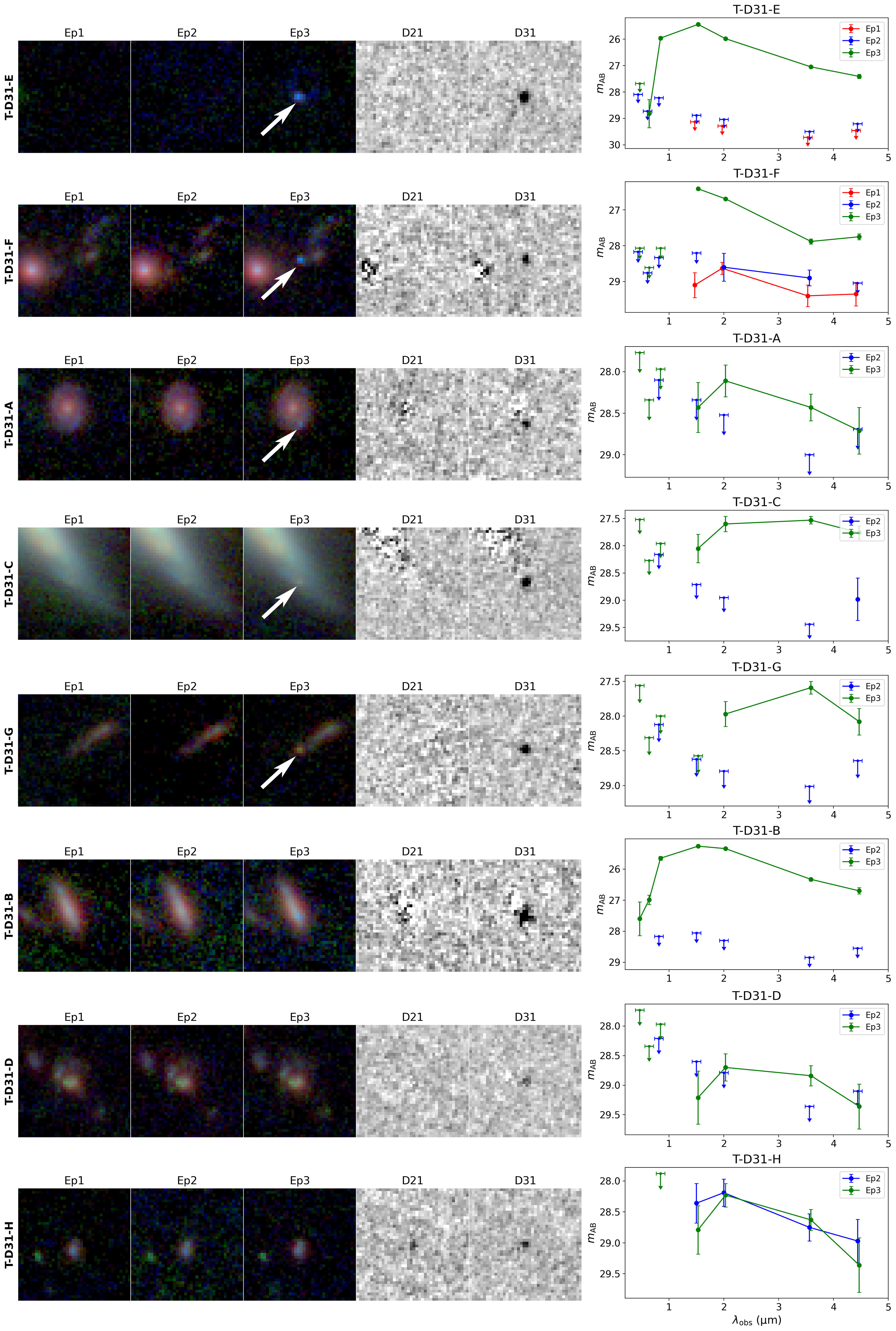}
    \caption{Same as Figure~\ref{fig:t-d31-all}, but for D31 transients. }
    \label{fig:t-d31-all}
\end{figure*}

\begin{figure*}
    \centering
    \includegraphics[width=\textwidth,height=\textheight,keepaspectratio]{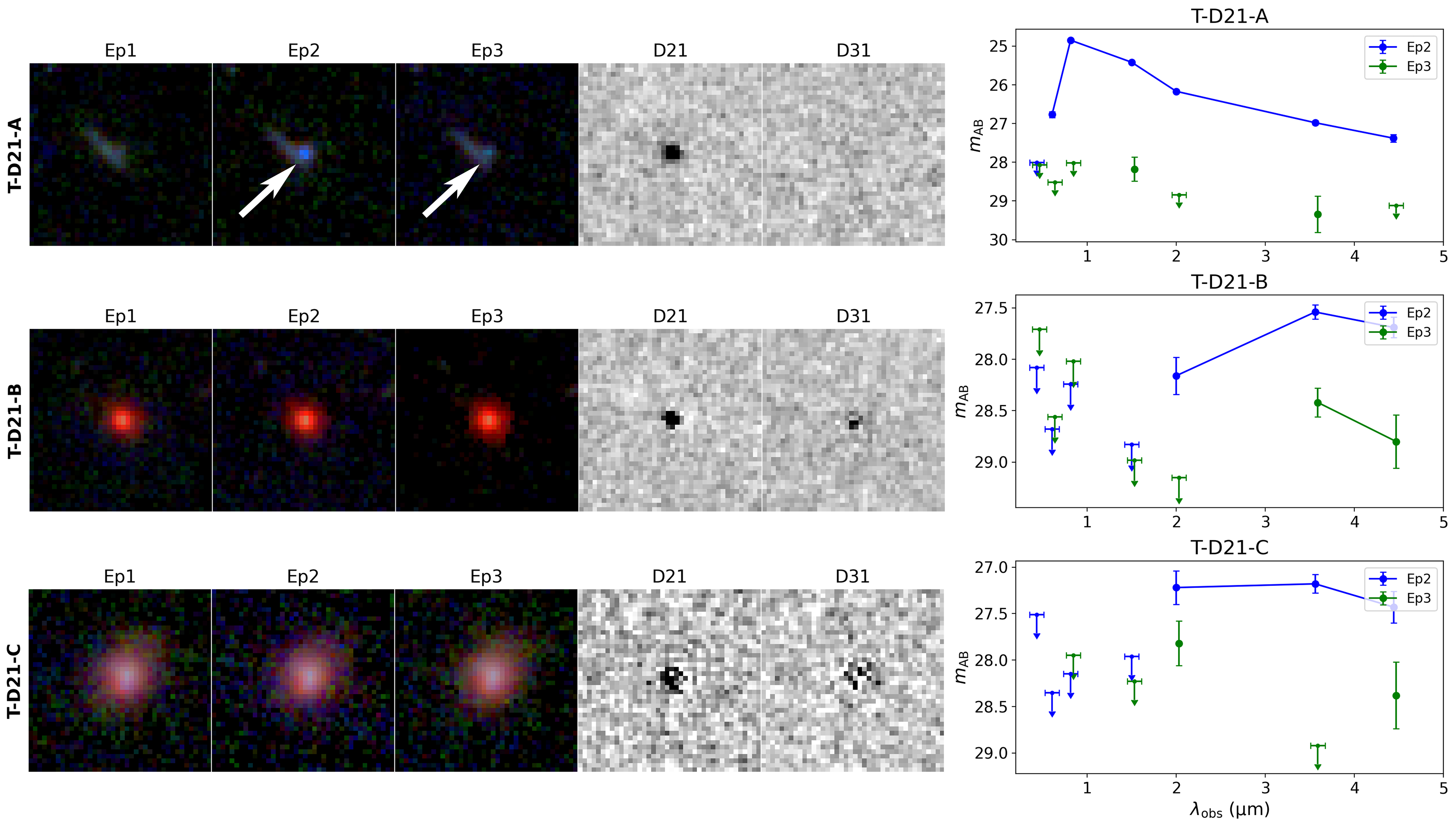}
    \caption{Same as Figure~\ref{fig:t-d31-all}, but for D21 (and D23) transients. }
    \label{fig:t-d21-all}
\end{figure*}

   Individual transients are described below, grouped by discovery epoch. 
Figures~\ref{fig:t-d13-all}, \ref{fig:t-d31-all}, and \ref{fig:t-d21-all} show 
their image stamps and SEDs. Quoted magnitudes have a superscript to indicate 
the corresponding epoch, e.g., $m_{356}^2$ stands for the magnitude in F356W in 
Ep2.
   
\subsection{D13 Transients}

\subsubsection{Hostless transients}

      (1) \texttt{T-D13-D:}\, The SED of this weak transient in Ep1 
peaked in F200W ($m_{200}^1=28.79\pm0.14$~mag) and dropped notably in F444W
($m_{444}^1=29.68\pm0.27$~mag). It decayed slowly, especially in 
the SW bands. Even in Ep3, it was still detected with ${\rm S/N}\approx 3.1$ in
F200W ($m_{200}^3=29.73\pm0.35$~mag). 

      (2)  \texttt{T-D13-K:}\, This transient's Ep1 SED also peaked at F200W  
($m_{200}^1=28.16\pm0.09$~mag) and dropped only slightly in F444W\null. The 
transient was barely visible in Ep2 and disappeared in Ep3. 

      (3) \texttt{T-D13-L:}\, This was a blue transient, with its Ep1 SED
peaking in F150W ($m_{150}^1=27.13\pm0.07$~mag) and decreasing moderately to
F444W ($m_{444}^1=27.69\pm0.09$~mag). The transient disappeared in Ep2. There 
are no Ep2 ACS images of this transient because the object was in an image gap.

\subsubsection{Outskirt transients}

      (1) \texttt{T-D13-C:}\, This blue transient appeared in the outskirts of
a dwarf galaxy at $z_{\rm ph}=0.04$. Its Ep1 SED peaked in F150W 
($m_{150}^1=26.21\pm0.07$~mag or $M_H=-10.01$~mag) and dropped significantly to 
the LW bands ($m_{356}^1=27.66\pm0.10$~mag and $m_{444}^1=27.52\pm0.12$~mag). 
The transient became very faint in Ep2, only barely visible in F150W
($m_{150}^2=27.98\pm0.34$~mag).

      (2) \texttt{T-D13-E:}\, This transient was in the outskirts of its
disk-shaped, red host galaxy, which is invisible in the ACS F435W band. Our SED 
fitting gives $z_{\rm ph}=2.6$. 
(The host is very close to that of \texttt{T-D13-F}.)
The Ep1 SED of this transient peaks in F200W ($m_{200}^1=26.97\pm0.07$~mag;
or $M_V \approx -18.32$~mag) and drops sharply to the LW bands, becoming 
invisible in F444W\null. There are no Ep3 ACS images of this transient because 
the object was in an image gap.

      (3) \texttt{T-D13-I:}\, This blue transient appeared at the tip of a
red disk galaxy. The galaxy seems to be made of two components in the ACS 
images, but the split into two is likely due to dust obscuration in the middle 
of the galaxy. We derived $z_{\rm ph}=2.6$ for the galaxy as a whole. The 
transient's Ep1 SED peaks at F150W ($m_{150}^1=27.33\pm0.11$~mag or 
$M_B\approx -17.96$~mag) and decreases monotonically to F444W 
($m_{444}^1=28.97\pm0.34$~mag). It decayed very slowly and kept nearly the same 
brightness in Ep2 except in F150W, in which it dropped significantly 
($m_{150}^2=28.02\pm0.25$~mag). This is one of only two transients 
(the other being \texttt{T-D31-H}) in our sample that maintained its brightness 
over two epochs.

\subsubsection{Deeply-embedded transients}

      (1) \texttt{T-D13-A:}\, The transient was in the core region of a very
compact galaxy. The galaxy is prominent in all bands except ACS F435W and
has $z_{\rm ph}=0.25$. This transient is among the faintest ones in 
our sample and is the only one whose discovery-epoch SED peaks in the LW bands
($m_{356}^1=28.50\pm0.16$ and $m_{444}^1=28.50\pm0.22$~mag; 
$M_K\approx -11.74$~mag).

      (2) \texttt{T-D13-B:}\, This transient was embedded in a merging
system that has $z_{\rm ph}=0.4$. Its Ep1 SED is nearly flat in the first
three NIRCam bands at around 28.45--28.66~mag ($M_J\approx -12.7$ and 
$M_H\approx -12.9$~mag) and drops to $m_{444}^1=29.16\pm0.34$~mag. After the 
transient declined in Ep2, it was only barely detected in the LW bands
($m_{356}^2=29.14\pm0.30$ and $m_{444}^2=29.10\pm0.37$~mag).

      (3) \texttt{T-D13-F:}\, This transient coincided with a bright knot at 
one side of a small galaxy that has $z_{\rm ph}=2.22$. The ``transient'' might 
even have been created by variability of the knot (e.g., the knot is dominated 
by an AGN) rather than being a true transient. However, the AGN possibility 
seems unlikely because the knot is off center. As shown in 
Figure~\ref{fig:sed_hosts}, the SED fitting of the host galaxy is the least 
satisfactory of all, largely because the blue emission in the ACS bands cannot 
be well fitted. The NIRCam D23 images show almost no signal at the transient 
location (except in F356W where there might be a very weak signal), which means 
that the event was largely gone in the IR by Ep2. There are no ACS D23 images 
for this transient or its host as they fell in the gap of the Ep3 ACS images.  
Therefore we cannot rule out the possibility that the transient might still 
have been present in Ep2 at visible wavelengths. If so, that might explain the 
strange host SED in ACS because we had to use the Ep2 ACS images to construct 
the host SED\null. The transient's Ep1 SED peaks in F150W 
($m_{150}^1=26.75\pm0.09$~mag; $M_B\approx -18.2$~mag) 
and drops slightly to the red wavelengths, where it remains relatively flat at 
27.2--27.6~mag.
      
      (4) \texttt{T-D13-J:}\, The host is an irregular galaxy at 
$z_{\rm}=0.34$ that can be separated into two components. The transient was on 
the southern component. The transient was significantly detected in Ep1 in 
F356W ($m_{356}^1=28.51\pm0.14$~mag) but was rather weak in other bands 
($M_K\approx -12.34$~mag by interpolating between F356W and F200W; 
$M_J\approx -12.45$~mag by using F150W). The transient vanished completely in 
Ep3.

\subsection{D31 Transients}

\subsubsection{Hostless transients}

      (1) \texttt{T-D31-E:}\, NIRSpec spectroscopy (Section 5.2) showed that 
the transient was a type~Ia supernova at $z=1.64$. This was one of the 
brightest and bluest transients in our sample with its Ep3 SED peaking in F150W
($m_{150}^3=25.44\pm0.02$~mag; $M_V\approx -18.94$~mag) and declining 
monotonically to F444W ($m_{444}^3=27.41\pm0.08$~mag). The transient was also 
prominent in the Ep3 ACS F814W ($m_I^3=25.96\pm0.06$~mag) but was only barely 
visible in F606W ($m_V^3=28.82\pm0.53$~mag) and completely invisible in 
F435W\null. It was invisible in any band in Ep1 or Ep2. 

     (2) \texttt{T-D31-F:}\, This was a blue transient with its Ep3 SED peaking 
in F150W ($m_{150}^3=26.41\pm0.03$~mag) and declining to the two LW bands at 
approximately the same brightness ($m_{356}^3=27.88\pm0.07$ and
$m_{444}^3=27.75\pm0.08$~mag). It was not detected in the ACS images, however.
This transient already emerged in NIRCam in Ep1 and kept nearly the same 
brightness through Ep2, especially in F200W ($m_{200}^1=28.63\pm0.17$ and 
$m_{200}^2=28.60\pm0.39$~mag).

\subsubsection{Outskirt transients}

      (1) \texttt{T-D31-A:}\, The host galaxy is a face-on spiral galaxy at
$z_{\rm ph}=2.6$, and the transient was located at the tail of one of its 
spiral arms. The transient's Ep3 SED peaks in F200W 
($m_{200}^3=28.11\pm0.19$~mag;  
$M_V\approx -17.18$~mag) and drops moderately to F444W 
($m_{444}^3=28.71\pm0.28$~mag); in the blue side, it was still detected in
F150W (although weak; $m_{150}^3=28.43\pm0.30$~mag) but was invisible in 
the ACS bands. Strangely, there is a weak hint that it might have appeared in 
Ep2, albeit only in ACS F606W.

      (2) \texttt{T-D31-C:}\, This transient appeared at the tip of a large 
disk galaxy, which has $z_{\rm ph}=0.21$. The transient's Ep3 SED has nearly 
the same brightness in F200W and F356W ($m_{200}^3=27.60\pm0.14$ and
$m_{356}^3=27.53\pm0.07$~mag; $M_H\approx -12.25$ and $M_K \approx -12.18$~mag) 
and drops off to both the blue and red sides. The transient was not detected in 
any of the ACS bands. There is a weak hint that it might already have been 
visible in Ep2, but only in F444W ($m_{444}^2=28.98\pm0.39$~mag).

      (3) \texttt{T-D31-G:}\, This transient appeared at the tip of a small
disk galaxy at $z_{\rm sp}=2.64$ (Section 5.2). The transient's Ep3 SED is 
similar to that of \texttt{T-D31-C}, peaking in F200W and F356W
($m_{200}^3=27.97\pm0.18$ and $m_{356}^3=27.59\pm0.09$~mag; 
$M_V\approx -17.35$ and $M_J \approx -17.73$~mag). 
However, there could be a complication: at the exact location of the transient,
there seems to be a weak detection in all Ep1 NIRCam images, 
suggesting that the transient might have gone off as early as in Ep1. For the 
photometry, however, we had to take the Ep1 images as the static
references for the difference images. Therefore, all Ep3 flux densities (based 
on the D31 difference images) could be underestimated.

\subsubsection{Deeply-embedded transients}

      (1) \texttt{T-D31-B:}\, The host galaxy has $z_{\rm ph}=0.16$, and its
morphology differs greatly over different wavelengths. In the NIRCam bands as 
well as the ACS F814W band, it is a regular disk galaxy, but the disk becomes 
asymmetric in the ACS F606W and even more so in F435W. The transient was 
located very close to the brightest core of the host in F606W\null. This core 
is very weak in F435W\null. The transient has a very blue Ep3 SED, peaking in 
F150W ($m_{150}^3=25.26\pm0.04$~mag;  $M_H \approx -14.0$~mag) and dropping 
monotonically to F444W ($m_{444}^3=26.70\pm0.10$~mag).

      (2) \texttt{T-D31-D:}\, The host is the major component of a merging
pair at $z_{\rm ph}=0.38$, and the transient was at its central region. The 
transient's Ep3 SED has a hump over F200W and F356W ($m_{200}^3=28.70\pm0.23$ 
and $m_{356}^3=28.84\pm0.17$; $M_H\approx -12.5$ and $M_K\approx -12.4$~mag) 
and drops by ${>}0.5$~mag on both sides. The transient was not detected in Ep2.

      (3) \texttt{T-D31-H:}\, The host is a small, compact galaxy at
$z_{\rm sp}=1.90$ (Section 5.2). While the transient happened at the edge of 
the host, they are severely blended, and therefore the transient is considered 
deeply embedded. Similar to \texttt{T-D13-I} discussed above (a declining 
transient), this rising transient kept its brightness nearly the same over two 
epochs. Its SED peaks in F200W in both epochs ($m_{200}^2=28.19\pm0.22$ 
and $m_{200}^3=28.23\pm0.19$~mag; $M_R\approx -16.5$~mag), and the time 
evolution is the most obvious in the bluest and the reddest bands. In F150W, 
the transient dropped from $m_{150}^2=28.36\pm0.32$ to $28.79\pm0.39$~mag, and 
in F444W, it dropped from $m_{444}^2=28.97\pm0.35$ to $29.36\pm0.44$~mag.

\subsection{D21 (and D23) Transients}

\subsubsection{Outskirt transients}

      (1) \texttt{T-D21-A:}\, The host galaxy appears to be a small, faint 
dwarf galaxy in the NIRCam bands and in ACS F814W but is almost invisible in
ACS F435W and F606W.  The transient occurred at the tip of the small 
$z_{\rm ph}=0.04$ host in Ep2 and was still visible in Ep3. The transient was 
very blue with its Ep2 SED peaking in F150W
($m_{150}^2=25.42\pm0.02$~mag; $M_H\approx -10.80$~mag) and monotonically 
declining to F444W ($m_{444}^2=27.38\pm0.10$~mag). Its overall brightness 
dropped by ${>}2$~mag in Ep3 ($m_{150}^3=28.18\pm0.31$~mag), becoming almost 
invisible in F444W.

\subsubsection{Deeply-embedded transients}

     (1) \texttt{T-D21-B:}\, This transient was in the nucleus of a compact, 
very red galaxy that is invisible in the three ACS bands and has
$z_{\rm ph}=2.46$. The Ep2 SED of the transient is also red, peaking 
in F356W $m_{356}^2=27.54\pm0.07$~mag (or $M_J\approx -17.65$~mag) and 
dropping to being invisible in F150W and bluer bands. The transient decayed by 
$\sim$0.7~mag in F356W by Ep3 ($m_{356}^3=28.42\pm0.014$~mag). Strangely, there 
was a weak detection in the F814W Ep3 image at the exact position of the 
transient; upon careful examination, we believe that this is a false positive 
peak because there are similar peaks around this location that exist only in 
this image, most likely due to imperfect rejection of defects.
    
     (2) \texttt{T-D21-C:}\, The host is a compact, early-type galaxy at
$z_{\rm ph}=2.22$. The transient was at the edge of the core region but was
still deeply embedded in the galaxy. The transient's Ep2 SED has a hump over 
F200W and F356W ($m_{200}^2=27.22\pm0.18$ and $m_{356}^2=27.18\pm0.10$~mag; 
$M_R$ and $M_J \approx -17.8$~mag). In Ep3, the transient was significantly 
detected only in F200W and F444W ($m_{200}^2=27.82\pm0.24$ and
$m_{444}^2=28.38\pm0.36$~mag).

\section{NIRSpec Spectroscopy of Three Transients}

    Three of the D31 transients were observed by NIRSpec through the JWST 
Cycle 2 director's discretionary time (DDT) program ID~4557 (PI H.\ Yan). The 
three targets were selected to explore different types of transient:
\texttt{T-D31-E} is hostless, \texttt{T-D31-G} is at the outskirts of an 
edge-on disk galaxy, and \texttt{T-D31-H} is one of the most peculiar 
transients in that it maintained its brightness over $\sim$6 months from Ep2 to 
Ep3. Due to a series of mishaps, the actual execution of the NIRSpec 
observations was unfortunately delayed by more than 2 months. In the end, the 
observations were carried out on 2023 October 30 (starting at  03:00 UT), 
nearly 4 months after Ep3. Despite that, the results proved useful.

\subsection{Observations and Data Reduction}

   The NIRSpec observations used the Micro-Shutter Assembly (MSA) to observe 
the three targets simultaneously and the PRISM/CLEAR disperser/filter 
combination to cover the full wavelength range of 0.6--5.3~$\mu$m. We used the 
three-shutter configuration, which results in slitlets of 
$0\farcs2\times1\farcs38$ in size. The PRISM disperser offers the highest 
sensitivity, albeit with a low spectral resolution $R\sim100$ at 
$\sim$3~$\mu$m. The observations used the NRSIR2 readout mode, and the spectra 
were taken in a sequence of three pointings (dithers), each containing a single 
integration of 16 groups. The resultant total exposure time was just over 10 
hrs. Each of the three pointings corresponded to an identical pattern of open 
and closed shutters on the MSA but taken at a different position on the 
detector to minimize damage from bad pixels and improve flat-fielding. The 
placement of sources in slitlets used the eMPT software
\footnote{\url{https://github.com/esdc-esac-esa-int/eMPT_v1}
\citep{Bonaventura2023}}, a stand-alone \texttt{Fortran} program that finds the 
optimal position on the sky taking into account the object priorities, the 
number of dithers, the presence of defective shutters on the MSA, and a search 
radius. The placement of the three transient targets defined the center and 
orientation of the MSA\null. In addition to the three transients, we observed 
as many ``filler'' objects (e.g., X-ray and radio sources) as possible. The 
mask configuration involved several steps---definition of priority classes of 
sources, the identification of reference stars, the placing and inspection of 
configured sources. In total, 28 sources were configured on the MSA.

    To reduce the spectra, we started from the Level 1b products retrieved from 
the MAST\null. We first processed them through the {\tt calwebb\_detector1} 
step of the standard JWST pipeline 
\citep[version 1.15.1; ][]{Bushouse24_jwppl} 
in the calibration context of {\tt jwst\_1299.pmap}. We then further processed 
the output ``rate.fits'' files through the {\sc msaexp} package 
\citep[version 0.8.5; ][]{Brammer_msaexp2023}, which provides an end-to-end 
reduction to the final spectral extraction. The steps in the calibration
stage included removing the $1/f$ noise pattern and the ``snowball'' defects, 
subtracting the bias level, applying flat-fielding and path-loss correction, 
and doing flux calibration. The background subtraction used the background
measurement in the nearest blank slit. In the final stage, {\sc msaexp} traced 
spectra on all individual exposures and combined all single exposures with 
outlier rejection.

   Aside from the three transients, discussed below, 14 of the 25 filler 
objects gave secure redshifts, which are presented in Appendix B. 

\subsection{Transient Identifications and Redshifts}

\begin{figure*}
    \centering
    \includegraphics[width=\textwidth,height=\textheight,keepaspectratio]{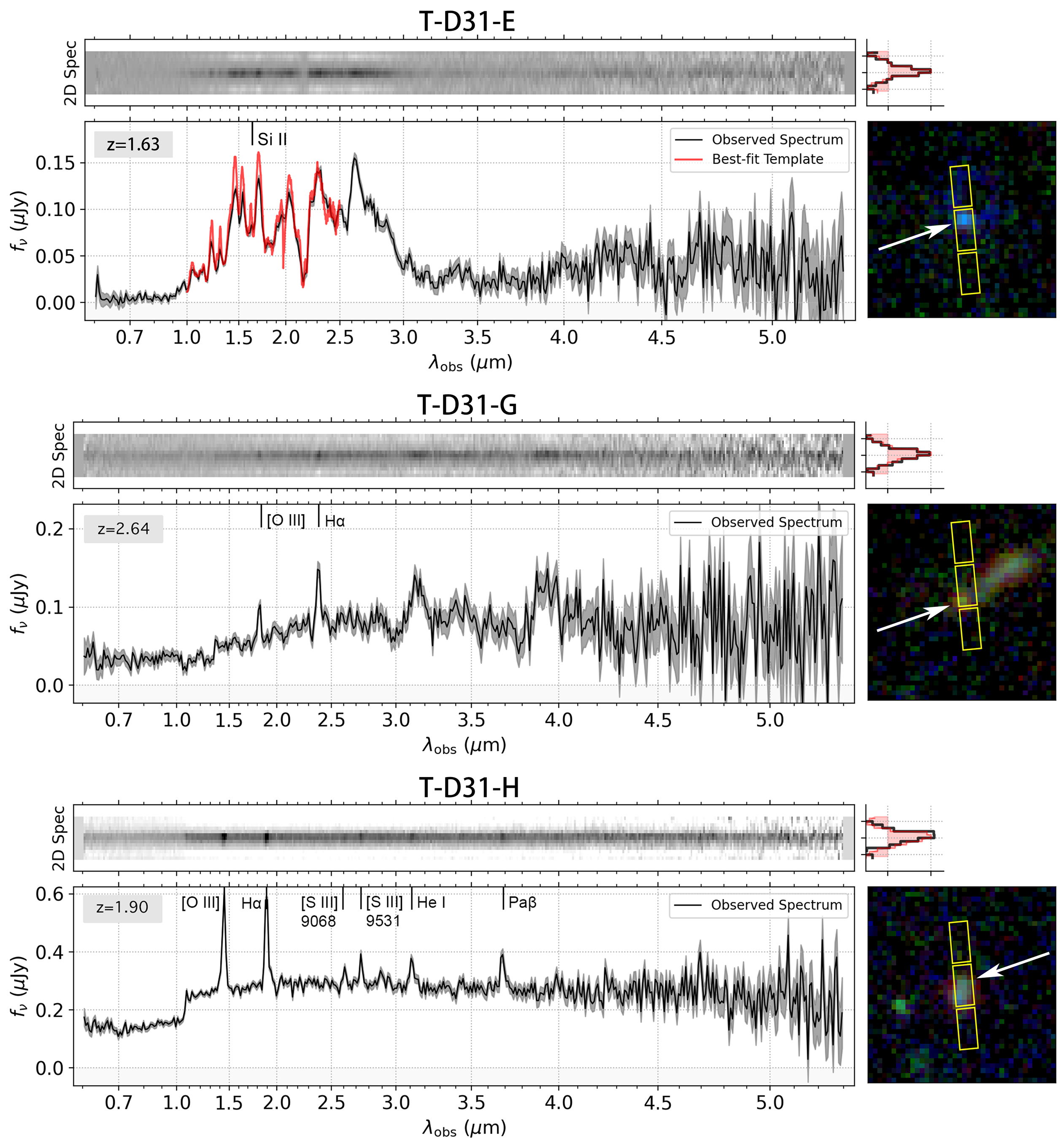}
    \raggedright
    \caption{JWST NIRSpec prism spectra of the three identified transients. 
    For each object, the reduced 2D and 1D spectra are shown in the top and 
    bottom panels, respectively. The gray area around the 1D spectrum indicates
    the 1$\sigma$ uncertainty range. On the right side of each block, the 
    three-shutter slit placement is overlaid on the Ep3 color image, which has 
    the same color scheme and field of view as in Figure~\ref{fig:t-d13-all}. 
    The location of the transient is indicated by an arrow in each.
    For \texttt{T-D31-E}, the superposed spectrum in red is the best-fit 
    template, which is SN~1999aa (an SN~Ia) redshifted to $z=1.64$ and evolved 
    to 33 rest-frame days after maximum.
    }
    \label{fig:spec_all}
\end{figure*}

  Figure~\ref{fig:spec_all} shows the 2D and 1D spectra of the three transient 
targets. A detailed analysis will be deferred to an upcoming paper 
(L. Wang et al., in preparation), and here, we only briefly summarize the 
results.

   Despite the delayed observations, the hostless transient  \texttt{T-D31-E}
was still bright enough to give a high-S/N spectrum. It matched an SNe~Ia 
spectra at $z=1.64$ that is $\sim$33 rest-frame days after maximum. One such 
fit is shown in the top panel of Figure~\ref{fig:spec_all} using SN~1999aa as 
the template. Using SN~2000cx as the template (not shown) gives a similar 
result. As the NIRCam Ep3 was 115.2 days (i.e., 43.6 rest-frame days) before 
the NIRSpec observations, Ep3 in fact caught this event $\sim$10.6 rest-frame 
days before maximum. 

   The two transients \texttt{T-D31-G} and \texttt{T-D31-H} had faded by the 
time of the NIRSpec observations, and their spectra are dominated by the host 
galaxies. Both galaxies are strong emitters of multiple lines, identified in 
Figure~\ref{fig:spec_all}. Secure redshifts are $z=2.64\pm0.01$ and 
$1.90\pm 0.01$, respectively.

\section{Discussion}

\subsection{Transient rate}

    Seeing 21 unique transients over a year in the 14.16~arcmin$^2$ area of the 
JWIDF implies an apparent transient rate of 
$\sim$1.48~events~arcmin$^{-2}$~yr$^{-1}$ at $m_{356}\la 29.0$~mag. The 
6-month cadence implies a rate of 
$\sim$0.78--0.92~events~arcmin$^{-2}$~6 months$^{-1}$, considering that there 
were 13 transients (10 in D13 and three in 
D21) and 11 transients (3 in D21 and 8 in D31) in the first and the 
second halves of the year, respectively. Had only two epochs spanning 1 yr 
(Ep1 and Ep3) been observed, the 18 transients detected would imply an apparent 
transient rate of $\sim$1.27~events~arcmin$^{-2}$~yr$^{-1}$. These estimates 
include no correction for survey incompleteness at faint magnitudes. 

    For comparison, in JADES data reaching $\sim$30.0~mag, 
\citet[][]{DeCoursey2025a} found 79 transients in 25~arcmin$^2$ with two epochs 
of data spanning over a year ($\sim$40 transients per epoch).
This translates to 3.16~events~arcmin$^{-2}$~yr$^{-1}$, $\sim$2$\times$ higher 
than ours. The difference is understandable because of their deeper data.
    
\begin{figure}
    \centering
    \includegraphics[width=0.41\textwidth,height=\textheight,keepaspectratio]{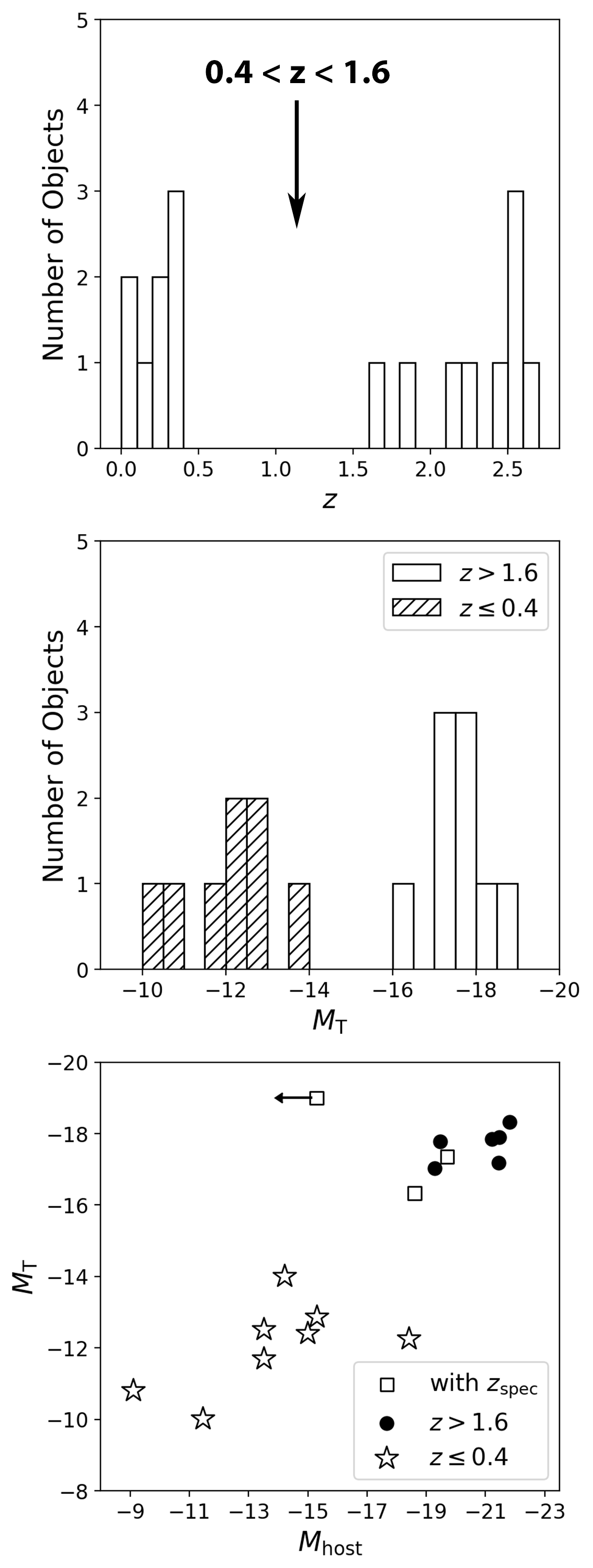}
    \raggedright
    \caption{Distributions of the transient redshifts (top; spectroscopic and
    photometric), the transient absolute magnitudes $M_{\rm T}$ (middle), and 
    the absolute magnitudes of the transient hosts $M_{\rm host}$ (bottom), 
    based on the values reported in Table~\ref{tab:trans_zM}. The point with 
    arrow shows the 2$\sigma$ limit for the undetected host of \texttt{T-D31-E}.
    }
    \label{fig:zM}
\end{figure}

\begin{table}
    \centering
    \scriptsize
    \caption{Redshifts and Absolute Magnitudes of the Transients and Their Hosts}
    \begin{tabular}{lllr}
    \hline\hline
    Transient ID & 
    \multicolumn{1}{c}{$z$} & 
    \multicolumn{1}{c}{$M_{\rm T}^w$} 
    & $M_{\rm host}^{V}$ \\[1ex] \hline 
     T-D13-A & 0.25 & $-11.7$ & $-13.5$ \\
     T-D13-B & 0.40 & $-12.9$ & $-15.3$ \\
     T-D13-C & 0.04 & $-10.0$ & $-11.5$ \\
     T-D13-J & 0.34 & $-12.5$ & $-13.5$ \\
     T-D31-B & 0.16 & $-14.0$ & $-14.2$ \\
     T-D31-C & 0.21 & $-12.3$ & $-18.4$ \\
     T-D31-D & 0.38 & $-12.4$ & $-15.0$ \\
     T-D21-A & 0.04 & $-10.8$ & $-9.1$ \\
\hline 
     T-D13-E & 2.60 & $-18.3$ & $-21.8$ \\
     T-D13-F & 2.20 & $-17.8$ & $-19.5$ \\
     T-D13-I & 2.60 & $-17.9$ & $-21.2$ \\
     T-D31-A & 2.60 & $-17.2$ & $-21.4$\\
     T-D31-E & 1.64$^*$ & $-18.9$ &  ${>}{-}15.3$ \\
     T-D31-G & 2.64$^*$ & $-17.4$ & $-19.7$ \\
     T-D31-H & 1.90$^*$ & $-16.3$ & $-18.6$ \\
     T-D21-B & 2.46 & $-17.0$ & $-19.3$ \\
     T-D21-C & 2.22 & $-17.8^{\dagger}$ & $-21.5$ \\
\hline
    \end{tabular}
    \raggedright
    \tablecomments{(1) The transients are separated by redshifts into the 
    ``low-$z$'' and ``mid-$z$'' groups by the horizontal line. The three 
    redshifts marked by $^*$ are spectroscopic (Section~5), and others are 
    photometric redshifts of the host galaxies (Section~3.3, 
    Figure~\ref{fig:sed_hosts}). The ``hostless'' transients are not included in 
    this table except for \texttt{T-D31-E}, which has $z_{\rm sp}$. 
    (2) $M_{\rm T}^w$ are the absolute magnitudes of the transients with $w=H$ 
    for the low-$z$ transients,  $w=V$ for mid-$z$ except for \texttt{T-D21-C}, 
    and $w=R$ for \texttt{T-D21-C} (marked by $^\dagger$).
    (3) $M_{\rm host}^V$ are the absolute magnitudes of the transient host
    galaxies in the rest-frame $V$ band.
    }
    \label{tab:trans_zM}
\end{table}

\subsection{Redshift and absolute magnitude distributions}

   Three of our transients have $z_{\rm sp}$ either from the transient itself 
or from the host galaxies (Section~5, Figure~\ref{fig:spec_all}), while 14 
others have $z_{\rm ph}$ from their hosts (Section~3, 
Figure~\ref{fig:sed_hosts}). The redshift distribution of these 17 transients 
is shown in Figure~\ref{fig:zM}. Somewhat surprisingly, the distribution shows 
a wide gap at $0.4<z<1.6$. A concern is whether this gap could be artificially 
created by some systematic errors in the $z_{\rm ph}$ derivations of the host 
galaxies. We investigated the $z_{\rm ph}$ accuracy, which is presented in 
Appendix~C. However, the result is inconclusive due to small number statistics 
in this stage. For the sake of the argument, here, we treat this gap as being 
real and separate the transients two groups: ``low-$z$'' ($z\leq 0.4$) and 
``mid-$z$'' ($1.6 \le z\la 2.6$).

   Absolute magnitudes corresponding to the peak of each transient's SED are 
reported in Section~4. For comparisons, it is more useful to calculate absolute 
magnitudes in the same rest-frame band. This is easy for the low-$z$ group, 
where rest-frame $H$-band (1.6~\micron) magnitudes (denoted $M_{\rm T}^{H}$)  
can be determined from $m_{150}$ and $m_{200}$. The situation is slightly more 
complicated for the mid-$z$ sources because of their redshift range and the 
wavelengths observed. The most useful choice is rest $V$ band (0.55~$\mu$m 
denoted $M_{\rm T}^{V}$), which for most mid-$z$ sources can be calculated from 
$m_{150}$ and $m_{200}$. \texttt{T-D21-C} has to be treated differently: the 
rest-frame $V$ band for this transient would be in F150W, which is a 
nondetection. Therefore, this source's absolute magnitude in rest-frame $R$ 
($M_{\rm T}^{R}$, 0.65~\micron) was calculated from $m_{200}$.
Table~\ref{tab:trans_zM} gives the $M_T$ values along with the redshifts, and 
Figure~\ref{fig:zM} shows their distribution. The low-$z$ and mid-$z$ groups 
segregate in $M_{\rm T}$ with a gap of $\sim$2~mag. The segregation is not 
because of the different rest-frame passbands in use, because the constraints 
from the ACS data on the D31 and D21 transients (no ACS data available for the 
D13 transients) in the low-$z$ group indicate that their $M^V$ would be even 
fainter than their $M^H$. In other words, the gap in the luminosity 
distribution is due to the gap in their redshifts: the transients in the 
low-$z$ group are less luminous than those in the mid-$z$ group. 

   The host-galaxy absolute magnitudes in the rest-frame $V$ band 
($M_{\rm host}^{V}$ are also reported in Table~\ref{tab:trans_zM}. The right 
panel of Figure~\ref{fig:zM} compares $M_{\rm T}$ and $M_{\rm host}$. While 
there seems to be a linear correlation between the two, a large part of that 
comes from the dependence of both luminosities on redshift. Limited statistics 
could also play a role as is hinted at by the two outliers, one being the 
hostless \texttt{T-D31-E}, which had a bright transient but faint host, and the 
other being \texttt{T-D31-C}, which had a faint transient but relatively bright 
host.

\subsection{Transients in the mid-$z$ group: candidate SNe}

    The mid-$z$ transients are at $z>1.6$ and have $M_{T}^V<-16.0$~mag. These 
absolute magnitudes are consistent with various types of SNe, and indeed,
\texttt{T-D31-E} was confirmed spectroscopically as an SN~Ia. We attempted to 
classify the other transients based on their SEDs and evolution (including SED 
upper limits in the epoch when a transient did not yet happen or had faded away:
Tables~\ref{tab:catD13}, \ref{tab:catD31}, and \ref{tab:catD21}) by checking 
the consistency with known SN types. Because we know precisely the time 
separations between epochs, the time evolution imposed a strong constraint on 
the fit. Due to the limited number of passbands in the SEDs of our transients, 
however, we did not expect an accurate SN type classification. The main purpose 
of the discussion here is to demonstrate how well the transients in this group 
are consistent with being SNe.

   We used the \texttt{SALT3NIR} \citep{Pierel2018,Pierel2022} spectral 
templates for SNe~Ia and the spectral templates of \cite{Gilliland1999} for 
Type IIP, IIL, Ib/c, and hypernovae. The approach is similar to that presented 
by \citet[][]{Yan2023c}. Basically, we fitted the SEDs (and the SED upper 
limits) to the templates by minimizing the $\chi^2$ with redshift priors using 
the values in Table~\ref{tab:trans_zM}. As these are mostly $z_{\rm ph}$, the 
redshift was allowed to vary slightly (within $\Delta z\pm 0.5$)
during the fit. The SEDs included only the NIRCam bands because these targets 
were not detected by ACS (or lack ACS observations) in their discovery epochs. 
The best-fit templates were checked against the ACS upper limits (when 
available) to ensure that they did not violate these limits. 

\begin{figure*}
    \centering
    \includegraphics[width=\textwidth,height=\textheight,keepaspectratio]{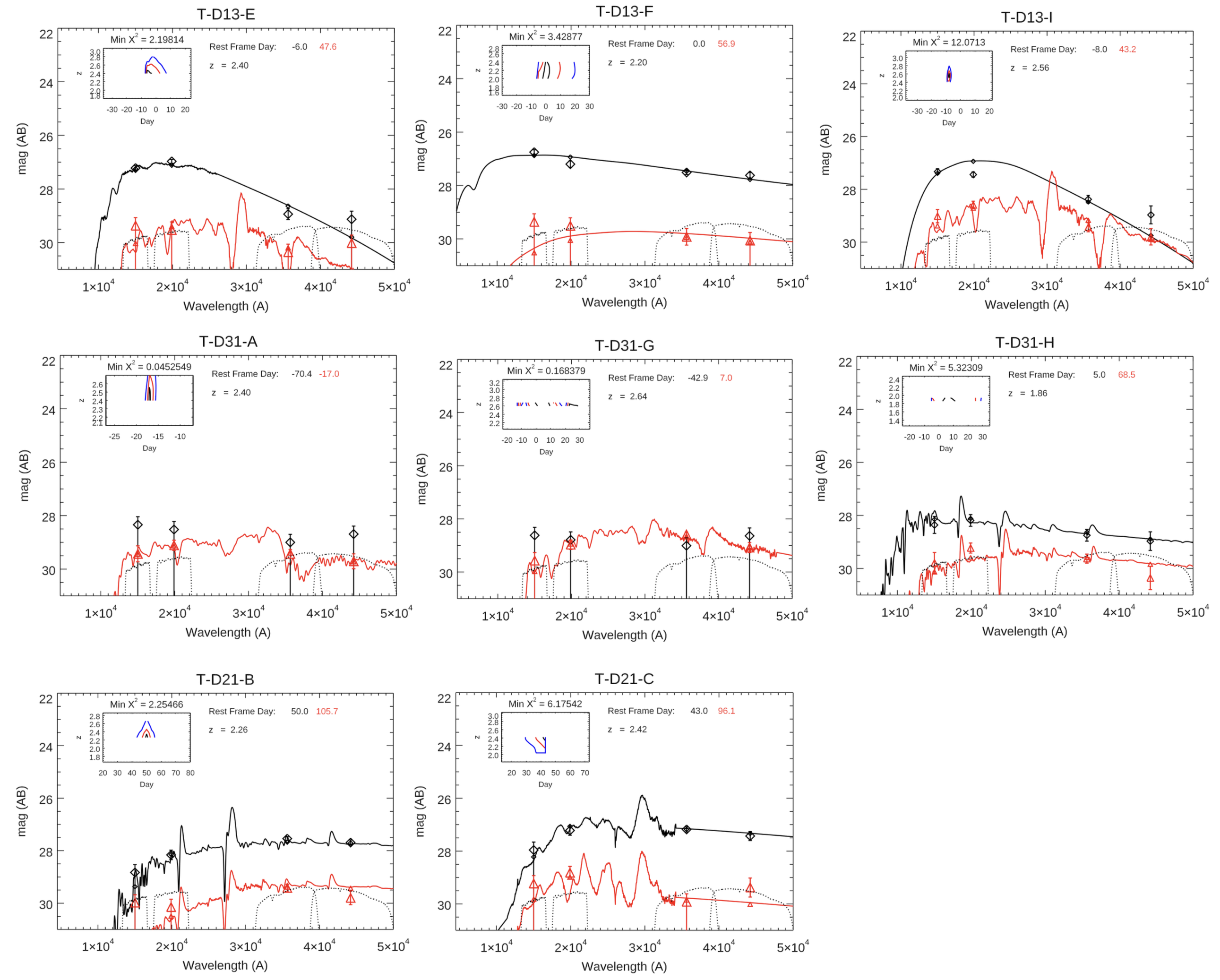}
    \raggedright
    \caption{Fitting to the NIRCam SEDs of the transients in the mid-$z$ 
    group using SN templates. The observed data are from 
    Tables~\ref{tab:catD13} to \ref{tab:catD21}, with the different epochs 
    (following the orders in these tables) shown in black and red. The latter
    (red) have an offset of 1~mag added for clarity of display. The
    curves are the best-fit SN templates, using the same color coding.
    Their corresponding rest-frame days with respect to the maximum are shown 
    (with the respective color) at the upper right of each panel, together with
    the fitted redshift. The insets show the confidence levels of the redshift 
    and epoch estimates, and the lines are the 1$\sigma$ (black), 2$\sigma$
    (red), and 3$\sigma$ (blue) contour levels. The minimum $\chi^2$ values are
    shown at the top of the insets. Note that \texttt{T-D31-A} and 
    \texttt{T-D31-G} do not have a template going through the black symbols; 
    this is because the libraries do not have templates on such early times 
    (70.4 and 42.9 rest-frame days as inferred, respectively) before the 
    maximum.
    }
    \label{fig:midzsne}
\end{figure*}

    Table~\ref{tab:midzsne} and Figure~\ref{fig:midzsne} summarize the 
SED-fitting results. Table~\ref{tab:midzsne} includes a quality flag for the SN 
type classification. Not surprisingly, many of them do not have high rankings 
due to the limited passbands. Nonetheless, the results at least show that these 
transients can be explained by SNe. The same arguments as in Section 6.1 give 
SN rates of 0.64~events~arcmin$^{-2}$~yr$^{-1}$ or 
0.39~events~arcmin$^{-2}$~6 months$^{-1}$. Again, these rates do not yet take 
into account the survey incompleteness nor the errors in $z_{\rm ph}$, 
and we will defer the full treatment to the forthcoming paper
(L. Wang et al., in preparation).
   
\begin{table}
    \centering
    \scriptsize
    \caption{Likely Supernova Types of the Mid-$z$ Transients}
    \begin{tabular}{lcccrrc}
Transient ID   & Type & $z$(SN) & $z$(host) & t1  & t2 & Qflag \\ 
\hline
T-D13-E & Ib/c & 2.40 & 2.60     & $-6.0$  &  47.6   & 2 \\
T-D13-F & IIn  & 2.20 & 2.20     &   0.0   &  56.9   & 2 \\
T-D13-I & Ib/c & 2.56 & 2.60     & $-8.0$  &  43.2   & 1 \\
T-D31-A & Ia   & 2.40 & 2.60     & $-70.4$ & $-17.0$ & 3 \\
T-D31-G & Ib/c & 2.64 & 2.64\rlap{$^*$} & $-42.9$ &  7.0    & 3  \\
T-D31-H & IIP  & 1.86 & 1.90\rlap{$^*$} &   5.0   &  68.5   & 1 \\
T-D21-B & IIP  & 2.26 & 2.46     &  50.0   & 105.7   & 2  \\
T-D21-C & hypn  & 2.42 & 2.22     &  43.0   &  96.1   & 1 \\
\hline 
    \end{tabular}
    \raggedright
    \tablecomments{(1) \texttt{T-D21-C} is best explained by a hypernova. 
    (2) $z$(SN) is the most probable redshift from the fit, while $z$(host)
    is the redshift from Table~\ref{tab:trans_zM}. 
    (3) The two values marked with ``*'' are spectroscopic redshifts (see 
    Section 5.2). The spectroscopic redshift of the host of \texttt{T-D31-H} is 
    $z=1.90$, which is slightly different from its $z$(SN) of 1.86 but is well 
    within the 1~$\sigma$ range (Figure~\ref{fig:midzsne}.
    (4) t1 and t2 are the times with respect to the maximum for the two epochs
    shown in Figure~\ref{fig:midzsne}, expressed in rest-frame days.
    (5) Qflag is the quality flag of the fit: ``3'' is reliable, ``2'' 
    is probable, and ``1'' is acceptable (the details will be presented in the
    forthcoming paper (L. Wang et al., in preparation).
    }
    \label{tab:midzsne}
\end{table}
     
\subsection{Transients in the low-$z$ group: gap transients?}

    There is no easy explanation for the low-$z$ transients, which are
at $z<0.4$ and have $-14.0 < M_{T}^H < -10.0$~mag. These luminosities are too 
high to be explained by classical novae, which have peak absolute $V$ 
magnitudes in the range of $M_V\approx -5$ to $-10$~mag 
\citep[e.g.,][]{Chomiuk2021}. The IR absolute magnitudes of novae are less well 
documented, but it is a general consensus that they have $V-K \approx 2$~mag 
around the peak \citep[e.g.,][]{BodeEvans2008} in the Vega system, which is 
$V-K\approx 0.2$~mag in the AB system. The most luminous nova reported in the 
literature, V1500 Cyg, has peak $M_H\approx -9.8$~mag (using the data from 
\citealt{Ennis1977} and adopting the distance of 2.0~kpc from 
\citealt{Ozdonmez2016}). Therefore, these transients are unlikely to be 
classical novae.

    SNe near maximum are far too luminous to explain the low-$z$ transients. 
SNe well past maximum could have the right absolute magnitudes, but the low-$z$ 
transients, at least the D21 and D31 sets, cannot be late-time SNe because they 
were not detected in the earlier epoch(s). There is no reason to believe that 
the low-$z$ D13 transients are different.

    Several tens of transients with luminosities populating the gap between 
those of SNe and classical novae were known in the pre-JWST era. These are
sometimes collectively called ``intermediate-luminosity optical transients'' or 
``gap transients'' and are thought to be related to high-mass stars and/or 
compact binaries 
\citep[e.g.,][and references therein]{Kasliwal2012, Soker2012, 
Pastorello2019a}. 
It is difficult to compare them to our transients because these pre-JWST gap 
transients usually lacked IR observations, and ours lack spectroscopic 
information. Nevertheless, known types of gap transients are worth considering.

\begin{figure*}
    \centering
    \includegraphics[width=\textwidth,height=\textheight,keepaspectratio]{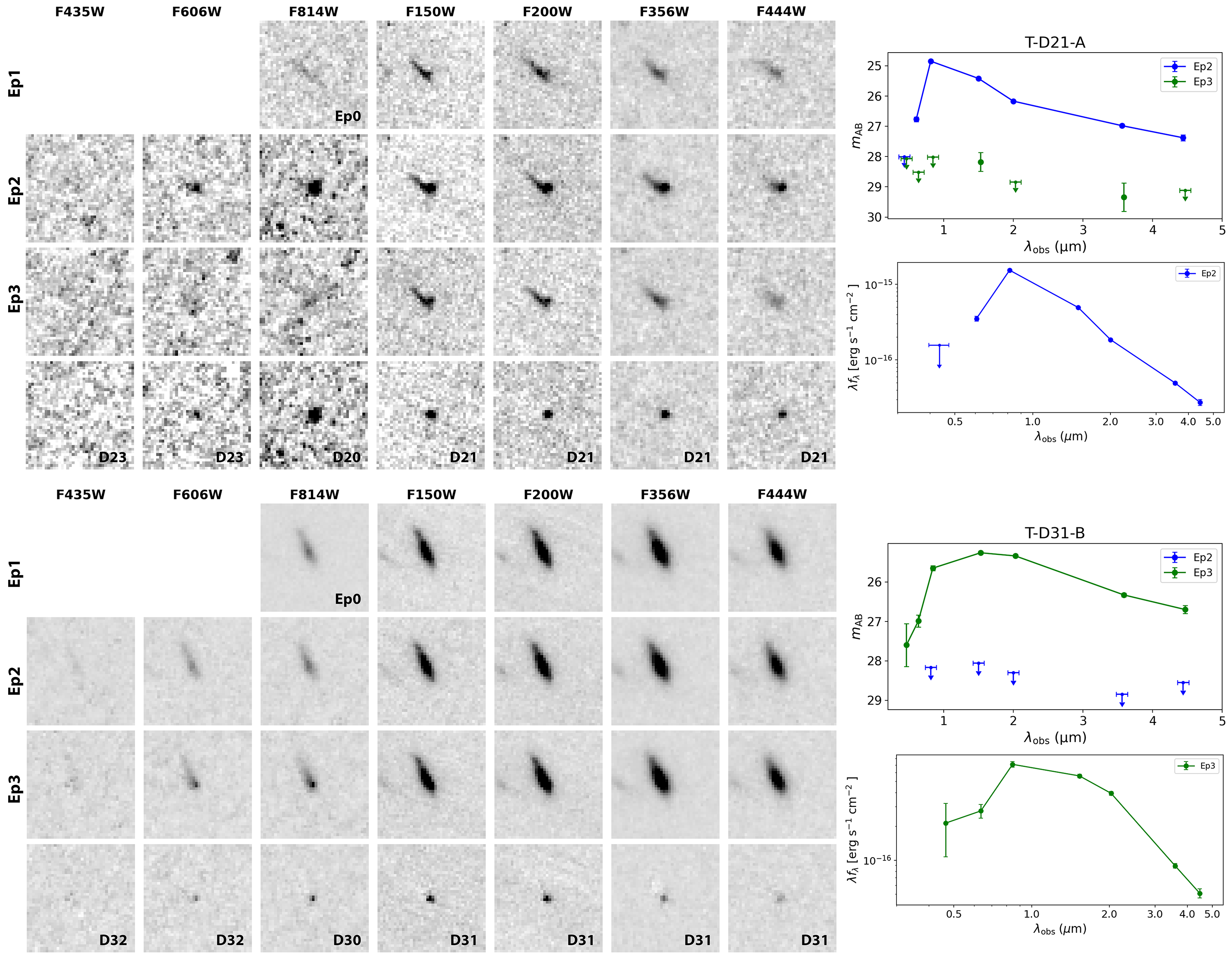}
    \raggedright
    \caption{Two examples of low-$z$ transients that could be SN impostors.
    The top and bottom panels show \texttt{T-D21-A} (host $z_{\rm ph}=0.04$) 
    and \texttt{T-D31-B} (host $z_{\rm ph}=0.16$), respectively, both of 
    which have ACS detections.
    Each object's negative images in the three epochs are shown 
    in the first three rows, and difference images from the most
    appropriate difference sets (as labeled) are shown in the last row. The 
    images are 2\farcs4$\times$2\farcs4 in size. There were no ACS 
    observations in Ep1, and in both cases, the F814W images in the Ep1 rows 
    are from the archival Ep0 data. The SED plots
    at right are in two forms: the top one is in AB magnitude versus
    wavelength and is the same as in Section~4, and the bottom is in
    $\lambda f_\lambda$ versus wavelength. (Both axes are logarithmic.)
    }
    \label{fig:snimpostor}
\end{figure*}

    A major type of gap transients are the so-called ``supernova impostors'' 
\citep{VanDyk2000}, which often show the characteristics of Type~IIn SNe 
(and hence the name). They are usually connected to the giant eruptions of 
luminous blue variables \citep[LBVs;][]{Humphreys1994}, which are 
$\gtrsim$10~\Msol\ supergiants becoming unstable in a late stage of life. The 
eruptions of LBVs can have a wide range of amplitudes and time scales
\citep[e.g.,][]{Smith2011LBV}, and some of those match the low-$z$ transients. 
Most of the known LBVs are blue in the visible range, and it is difficult to 
judge whether our objects have similar SEDs. Most of the D21 and D31 transients 
that have ACS photometry have only upper limits in the three bands. 
\texttt{T-D21-A} and \texttt{T-D31-B} were significantly detected in the ACS, 
and Figure~\ref{fig:snimpostor} shows their per-epoch image stamps and SEDs. 
While neither transient is very blue, some known LBV outbursts have similar 
SEDs. One example is HR Carinae \citep[][their Figure 4(c)]{Humphreys1994}. 
Therefore, SN impostors could account for some of the low-$z$ transients.

   Another major type of gap transients is luminous red novae 
\citep[LRNe; e.g.,][]{Kulkarni2007}, which are believed to happen during the 
short common-envelope evolution phase of binaries 
\citep[e.g.,][and references therein]{Howitt2020}. The typical light curves of 
LRNe have two peaks separated by a few months. During the first peak, LRNe are 
dominated by emission in the $V$ band and bluer; during the second peak, which 
can stretch over $\sim$100 days, they become very red. Only a few LRNe have 
extensive IR observations, and AT~2017jfs in NGC 4470 has the most 
comprehensive visible-to-near-IR data over $\sim$1~yr 
\citep[][their Table A.1]{Pastorello2019c}. This object had an SED 
monotonically decreasing (in the AB system) from $J$ to $K$ band, consistent 
with the SEDs of most of the low-$z$ transients. In other words, our objects 
could be explained by LRNe in their second peak.

    A final possibility is kilonovae, which normally are not included in the 
gap transients. Kilonovae are caused by binary neutron star mergers 
\citep[][]{LLX1998, Metzger2010} and are $\sim$1000 times more luminous than 
classical novae; they could be in the luminosity range of our transients if 
discovered after the peak. AT~2017gfo in NGC~4993
\citep[][and the references therein]{Abbott2017_followup}, the first confirmed 
kilonova, was the afterglow of the famous gravitational-wave event GW170817 
\citep[][]{Abbott2017_discovery}. The source dropped from its peak of 
$H\approx 17.5$~mag to $H\approx 20.0$~mag (AB) in $\sim$10~days 
\citep[e.g.,][]{Cowperthwaite2017}. That was $M_H\approx -15.5$~mag to 
$M_H\approx -13$~mag based on the distance modulus of 33.0~mag from 
\citet{Cantiello2018}. However, AT~2017gfo became very red and had a 
monotonically increasing SED (in $f_\nu$ or AB) from $u$ to $K$ a few days 
after the initial burst \citep[see Figure 1 of][]{Cowperthwaite2017}, which is 
inconsistent with most of our objects.

   In summary, the low-$z$ transients could be similar to the gap transients.
However, this interpretation hinges on the assumption that they are indeed at
$z<0.4$ as their host galaxies' $z_{\rm ph}$ suggest.

\section{Summary}

   PEARLS three-epoch (6-month cadence), four-band NIRCam imaging data have 
revealed 21 transients to $m_{356}\la 29.0$~mag in 14.16~arcmin$^2$. A separate 
HST provided ACS observations of the same field contemporaneous with the last 
two NIRCam epochs. Among the 21 transients, 10 were found in Ep1, 8 were 
found in Ep3, and 3 were found in Ep2 and would have been missed had there 
not been such an epoch. The overall transient rate is 
$\sim$0.78--0.92~events~arcmin$^{-2}$~6 months$^{-1}$. In terms of the 
relations with their host galaxies, 5 transients were ``hostless'', 7 
were in the outskirt regions of their hosts and can be discerned directly in 
the images taken in the discovery epoch, and nine were deeply embedded in their 
hosts and can only be revealed by subtracting the host. 

    Low-resolution prism spectroscopy of three transients confirmed one 
hostless transient to be an SN~Ia at $z_{\rm sp}=1.63$. The other two 
transients were not detected, presumably because they had already faded away by 
the time of the NIRSpec observations ($\sim$4 months after Ep3). Nevertheless, 
the observations measured host-galaxy redshifts $z_{\rm sp}=2.64$ and 1.90. 

    Photometric redshifts based on the NIRCam and ACS photometry, together with 
the three spectroscopic redshifts, seem to suggest that these transients fall 
into two groups: a mid-$z$ group at $z>1.6$ and a low-$z$ group at $z<0.4$.
The groups are separated by a gap at $0.4<z<1.6$. For this reason, the absolute 
magnitudes of the transients are also separated into two groups: the mid-$z$ 
group have $M^V_{\rm T}\lesssim -16.0$~mag, and those in the low-$z$ group have 
$M^H_{\rm T}\gtrsim -14.0$~mag. The SEDs of the mid-$z$ transients are 
consistent with various types of SNe. The low-$z$ transients, on the other 
hand, are less easily explained. Their absolute magnitudes would make them 
broadly consistent with the so-called ``gap transients,'' which are much more 
luminous than classical novae but less luminous than SNe. Due to possible 
catastrophic failures that could bias the photometric redshifts to low-$z$, 
however, the interpretation of their being gap transients is only tentative at 
this point. Spectroscopy of their hosts will be needed for confirmation. If 
they are indeed at $z<0.4$, such transients will be interesting to investigate 
further. While it is possible that they are a mixture of different types of gap 
transients (e.g., SN impostors and LRNe), it is also possible that they 
contain some new kind(s) of transients. Prompt NIRSpec spectroscopy and 
long-term NIRCam monitoring with high cadences will be critical in order to pin 
down their true nature.

\begin{acknowledgements}

This work is based on the observations made with the NASA/ESA/CSA James Webb
Space Telescope (PIDs 1176, 2738, 4557) and Hubble Space Telescope (PID 17154)
and obtained from the Mikulski Archive for Space Telescopes,
which is a collaboration between the Space Telescope Science Institute 
(STScI/NASA), the Space Telescope European Coordinating Facility (ST-ECF/ESA),
and the Canadian Astronomy Data Centre (CADC/NRC/CSA). 

We thank the anonymous referee for the constructive comments that have
improved the quality of this paper.
H.Y. and B.S. acknowledge the support from the University of Missouri Research 
Council grant URC-23-029, JWST PID 4557, HST PID 17154, HST PID 16621 and
National Science Foundation grant No.\ AST-2307447.
Z.M. is supported by the NSF grants Nos. 1636621, 2034318, and 2307448.
L.W. is grateful to the NSF for the grant AST 1813825 and the STScI for the 
grant JWST-AR-05965, which partially supported his research.
C.N.A.W acknowledges funding from JWST PID 4557 and the  JWST/NIRCam contract 
to the University of Arizona NAS5-02015.
R.A.W., S.H.C., and R.A.J. acknowledge support from NASA JWST Interdisciplinary
Scientist grants NAG5-12460, NNX14AN10G, and 80NSSC18K0200 from GSFC.
J.F.B. was supported by National Science Foundation grant No. PHY-2310018.
J.-S.H. and C.C. are supported by Chinese Academy of Sciences South America 
Center for Astronomy (CASSACA) Key Research Project E52H540301. 
C.C. also acknowledges the support from the CMS grant CMS-CSST-2025-A07 and 
the NSFC grants Nos. 11803044 \& 12173045.
C.J.C. and N.J.A acknowledge support from the ERC Advanced Investigator Grant 
EPOCHS (788113). L.H. acknowledges support from the STScI grant JWST-AR-05965.
H.B.H. and S.N.M. acknowledge support from NASA JWST Interdisciplinary 
Scientist grant 21-SMDSS21-0013.

{The JWST and HST data presented in this article were obtained from the 
Mikulski Archive for Space Telescopes (MAST) 
at the Space Telescope Science Institute. 
The specific observations analyzed can be accessed via 
\dataset[doi:10.17909/ez4h-r505]{https://dx.doi.org/10.17909/ez4h-r505}. 
}
 
\end{acknowledgements}


\appendix
\counterwithin{figure}{section}
\counterwithin{table}{section}

\section{Light curves and Piecewise power-law SEDs of the Transients}

\begin{figure*}
    \centering
    \includegraphics[width=\textwidth,height=\textheight,keepaspectratio]{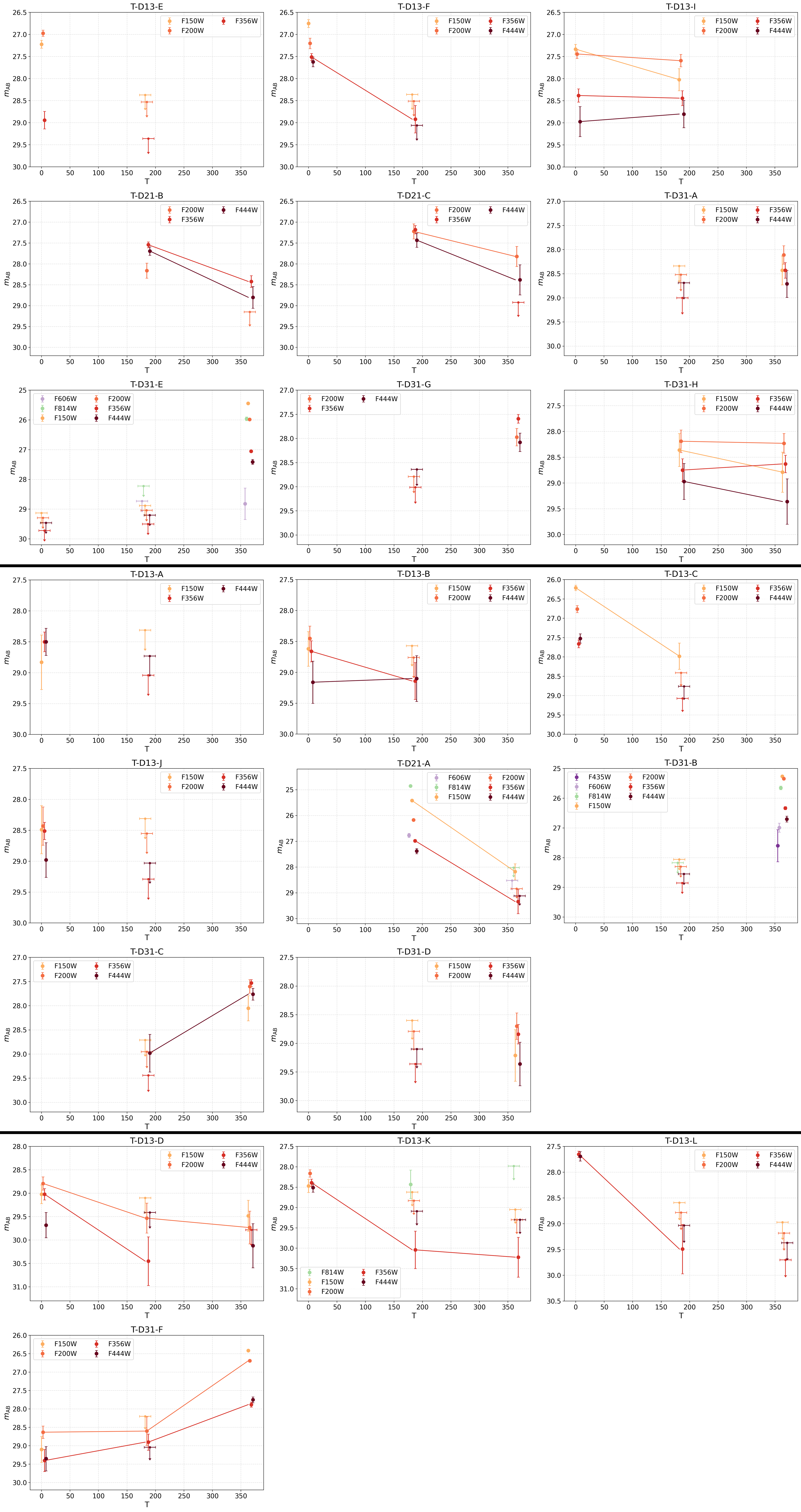}
    \raggedright
    \caption{Light curves of the transients. The passbands are indicated in
    the legends. The dates are {\rm MJD - 59765} days. 
    The transients are shown in three blocks based on redshift and separated by horizontal lines: mid-$z$ (top), low-$z$ (middle), no redshift (bottom). 
    }
    \label{fig:lc}
\end{figure*}

    For the sake of completeness, the light curves of our transients are shown 
in Figure~\ref{fig:lc}. 

\begin{figure*}
    \centering
    \includegraphics[width=\textwidth,height=\textheight,keepaspectratio]{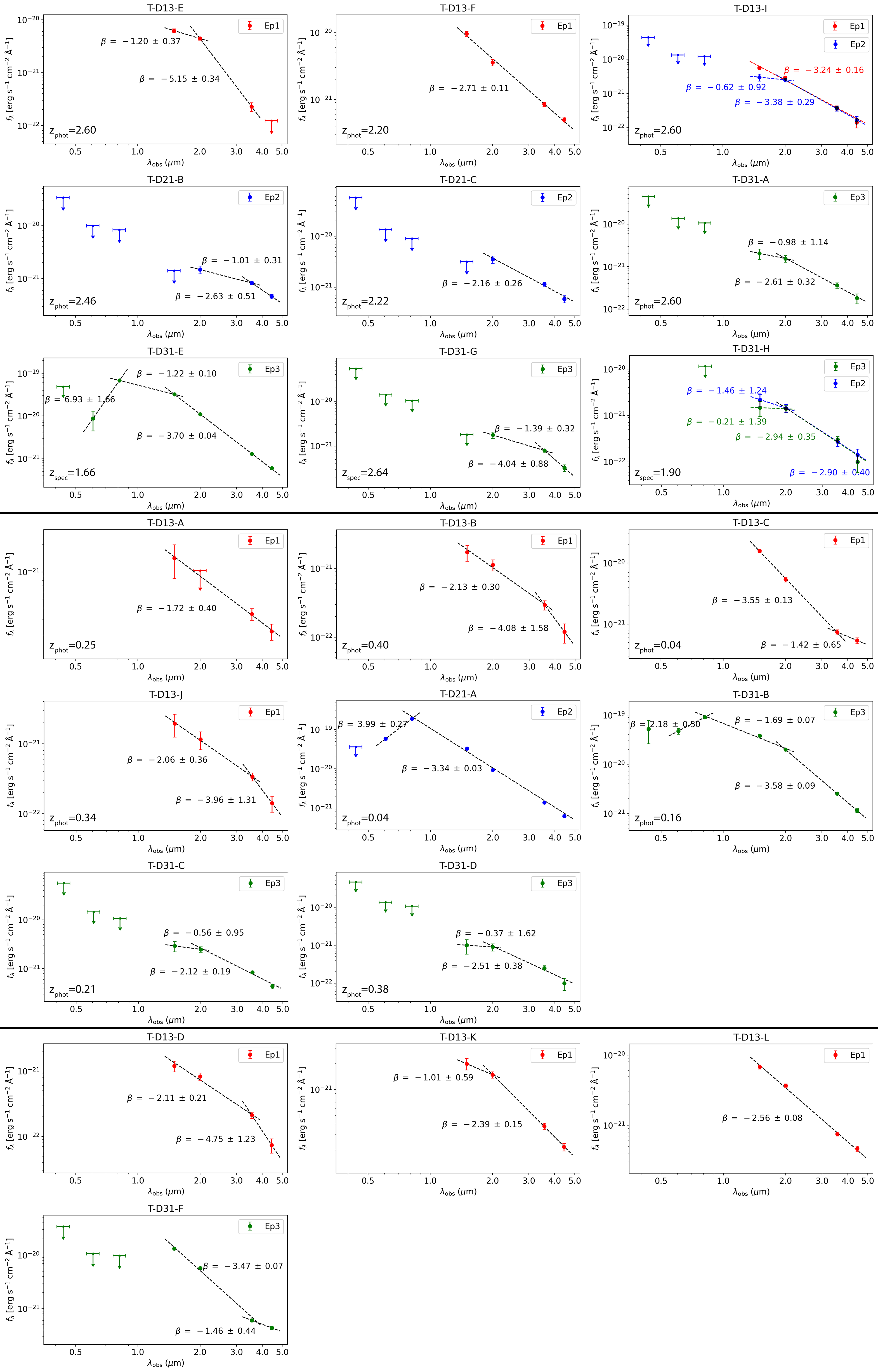}
    \raggedright
    \caption{Piecewise power-law fits (dashed lines) to the transient SEDs 
    (colored symbols). 
    The transients are grouped in three blocks as in Figure~\ref{fig:lc}.
    The SEDs are for each transient's peak epoch and expressed in $f_\lambda$ versus $\lambda$. Both axes are logarithmic. 
    The redshifts ($z_{\rm ph}$ or $z_{\rm sp}$) and indices $\beta$ for each power-law segment are labeled.
    }
    \label{fig:flam_sed_beta}
\end{figure*}

    The 21 transients all have similar SEDs regardless of whether they are in 
the mid-$z$ or low-$z$ groups. If expressed in $f_\lambda$, they all have 
decreasing SEDs in the four NIRCam bands from F150W to F444W\null. Those that 
were detected in the ACS bands all peak in F814W\null. This is shown in
Figure~\ref{fig:flam_sed_beta}. As it turns out, their SEDs can be well 
described by piecewise power laws in the form of 
$f_\lambda \propto \lambda^\beta$.

\newpage

\section{Filler Redshifts from DDT 4357}

   As mentioned in Section 5, the NIRSpec MSA follow-up spectroscopy observed 
25 filler objects in addition to the three transient targets. The table below 
summarizes the results. 

\begin{table*}[hbt!]
    \centering
    \caption{Coordinates, magnitude in F356W, and spectroscopic redshift for fillers from DDT 4557}
    \begin{tabular}{cccccc}
        NIRSpec ID & R.A. (deg) & Decl. (deg) & $m_{356}$ & $z_{\rm spec}$ & $Q$ \\ \hline 
        100001 & 265.0076266 & 68.9826853 & $27.80\pm0.07$ & \no & 0 \\ 
        200313 & 265.0081936 & 68.9749468 & $28.05\pm0.08$ & $2.08\pm0.01$ & 3 \\ 
        200326 & 265.1505180 & 68.9760296 & $27.46\pm0.05$ & \no & 0 \\ 
        200342 & 265.0392402 & 68.9776345 & $26.24\pm0.03$ & $1.88\pm0.01$ & 3 \\ 
        200351 & 265.0599834 & 68.9787689 & $21.66\pm0.01$\rlap{$^{\dagger}$} & $1.80\pm0.01$ & 3 \\ 
        200374 & 265.1420808 & 68.9811279 & $27.35\pm0.06$ & \no & 0 \\ 
        200382 & 265.1537943 & 68.9818101 & $28.49\pm0.12$ & $1.58\pm0.01$ & 3 \\ 
        200389 & 265.0061607 & 68.9819667 & $27.43\pm0.05$ & $3.95\pm0.01$ & 3 \\ 
        200404 & 265.0439362 & 68.9832430 & $27.17\pm0.06$ & $1.29\pm0.03$ & 1 \\ 
        200412 & 265.0656539 & 68.9840020 & $25.67\pm0.02$ & $1.90\pm0.02$ & 2 \\ 
        200424 & 265.0040064 & 68.9845482 & $28.71\pm0.12$ & \no & 0 \\ 
        200439 & 265.0576138 & 68.9855250 & $28.30\pm0.09$ & $1.94\pm0.01$ & 3 \\ 
        200445 & 265.0003883 & 68.9859689 & $27.73\pm0.07$ & $1.78\pm0.02$ & 2 \\ 
        200475 & 265.1385620 & 68.9889785 & $27.60\pm0.05$ & $0.01\pm0.03$ & 1 \\ 
        200476 & 265.0707513 & 68.9890466 & $27.84\pm0.05$ & \no & 0 \\ 
        200485 & 265.0710791 & 68.9898678 & $27.35\pm0.05$ & $1.65\pm0.02$ & 1 \\ 
        200506 & 265.0664248 & 68.9931117 & $28.46\pm0.08$ & \no & 0 \\ 
        200527 & 265.0621749 & 68.9959781 & $27.09\pm0.07$ & \no & 0 \\ 
        200556 & 265.1629469 & 68.9869338 & $20.37\pm0.01$ & $1.62\pm0.01$ & 3 \\ 
        500214 & 265.0631095 & 69.0114227 & \no & $0.87\pm0.01$ & 3 \\ 
        600060 & 265.1505550 & 68.9735036 & $22.41\pm0.01$ & $4.20\pm0.01$ & 3 \\ 
        600062 & 265.0628595 & 68.9737770 & $24.78\pm0.01$ & $1.82\pm0.02$ & 3 \\ 
        600115 & 265.0618379 & 68.9865172 & $24.97\pm0.01$ & $2.75\pm0.01$ & 3 \\ 
        600132 & 265.1408802 & 68.9913438 & $24.94\pm0.01$ & \no & 0 \\ 
        600171 & 265.1271301 & 69.0094838 & \no & $0.93\pm0.02$ & 3 \\ 
        \hline 
    \end{tabular}
    \raggedright
    \tablecomments{Coordinates are in the GAIA DR3 system. 
    Magnitudes in the F356W passband were measured on the epochs 1+2+3 stack. Redshift quality $Q$ is as follows.
    3, determined with at least two high emission lines with $S/N\ge5$; 
    2, determined with at least two emission lines at lower S/N; 
    1, very weak or no emission lines, determined using other spectral features such as Balmer break; 
    0: no emission lines or spectral features can be discerned.
    $Q=3$ $z_{\rm sp}$ values are solid, and $Q=2$ should be reliable. 
    Objects \#500214 and \#600171 fall outside of the JWIDF field, 
    and no $m_{356}$ can be determined. 
    The slit position of \#200351 (labeled with $\dagger$) 
    falls on the edge of a spiral galaxy, 
    whose F356W magnitude is quoted. 
    }
    \label{tab:specz_all}
\end{table*}

\section{Reliability of $z_{\rm ph}$}

    As mentioned in Section 6.2, the gap in the redshift distribution
(and hence the gap in the absolute magnitude distribution; see 
Figure~\ref{fig:zM}) is somewhat surprising, which leads to a concern whether it
is real or is artificially created due to erroneous $z_{\rm ph}$ 
derivations. Considering that there is a wide gap between the reddest ACS band 
(F814W) and the bluest NIRCam band (F150W), some catastrophic failures in 
$z_{\rm ph}$ are not unlikely. Therefore, we use the $Q=3$ $z_{\rm sp}$ 
presented in Appendix~B to investigate this problem. 

    Unfortunately, only a small number of these $Q=3$ $z_{\rm sp}$ are suitable 
for our purpose. When selecting the fillers for the NIRSpec MSA observations, 
we intentionally favored those that are weak or even undetected in the ACS 
bands, which means that most of the objects in Table~\ref{tab:specz_all} have 
detections in only four (in NIRCam) to five bands (four NIRCam bands plus the 
ACS F814W). Estimates of $z_{\rm ph}$ based on SEDs constructed using such a 
limited number of bands with detections are often unreliable. On the other 
hand, most of the objects in the low-$z$ group have detections in all seven 
bands. Therefore, we only use the $Q=3$ objects in Table~\ref{tab:specz_all} 
that have detections in seven bands for this investigation. 

\begin{figure*}
    \centering
    \includegraphics[width=\textwidth]{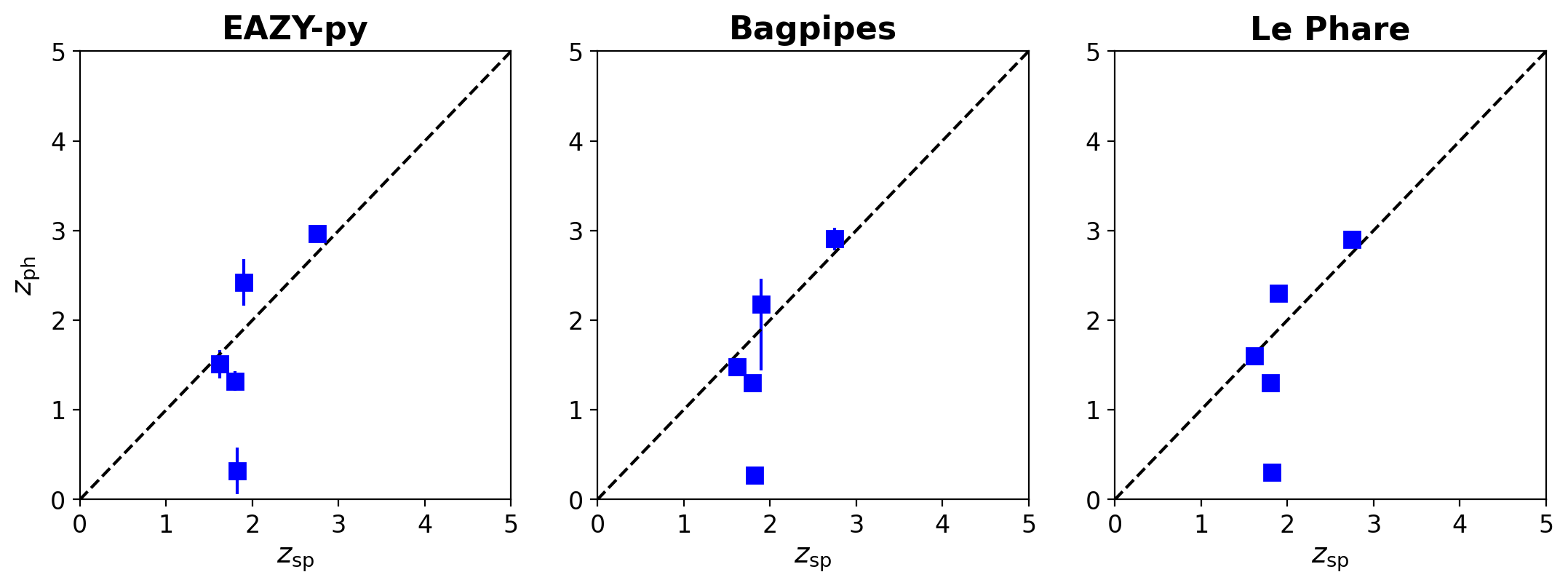}
    \raggedright
    \caption{Comparison of $z_{\rm ph}$ and $z_{\rm sp}$, using five $Q=3$ 
    objects in Table~\ref{tab:specz_all} that also have detections in all seven 
    bands (ACS and NIRCam). Three different SED-fitting methods are used to 
    derive $z_{\rm ph}$. There is one object suffering from a catastrophic 
    failure consistently in all three methods, which has $z_{sp}=1.82$ but with 
    $z_{\rm ph}=0.27$--0.32.
    }
    \label{fig:zph_zsp}
\end{figure*}

    There are only five objects that meet this requirement. In addition to
\texttt{EAZY-py}, we also use \texttt{Bagpipes} \citep[][]{Carnall2018} and
\texttt{Le Phare} \citep[][]{Arnouts1999, Ilbert2006} to derive $z_{\rm ph}$. 
The comparison of $z_{\rm ph}$ and $z_{\rm sp}$ is shown in 
Figure~\ref{fig:zph_zsp}.
One object, which has $z_{sp}=1.82$, always has a catastrophic failure in all
three methods with $z_{\rm ph}=0.27$--0.32. This indicates that the low-$z$ 
solutions ($z_{\rm ph}<0.4$) that we obtained in Section 3.3 could suffer from 
erroneous $z_{\rm ph}$. However, more $z_{\rm sp}$ would be needed to assess 
the fraction of such failures. On the other hand, the four mid-$z$ objects have
their $z_{\rm ph}$ in good agreement with $z_{\rm sp}$, suggesting that the
interpretation of the transients in the mid-$z$ group does not suffer from such
a problem.

\end{document}